\documentclass[10pt,letterpaper]{aprielvoice}
\usepackage[utf8]{inputenc}
\usepackage{lmodern}
\usepackage{amsmath,amssymb,mathtools}
\usepackage{array}
\usepackage{float}
\usepackage{xurl}
\usepackage{enumitem}
\usepackage{nameref}
\usepackage[noabbrev,nameinlink,capitalize]{cleveref}
\makeatletter
\patchcmd{\Gread@pdf}{\Gread@transgrouptrue}{\Gread@transgroupfalse}{}{}
\makeatother
\definecolor{linkblue}{HTML}{165A78}
\hypersetup{colorlinks=true,linkcolor=linkblue,citecolor=linkblue,urlcolor=linkblue,
  pdftitle={Apriel Voice 27B Preview: An Open Model for Speech, Reasoning, and Tool Use},
  pdfauthor={ServiceNow},
  pdfsubject={Speech reasoning, conversational tool use, multimodal adaptation, and efficient online distillation}}
\crefname{section}{Sec.}{Secs.}
\Crefname{section}{Section}{Sections}
\crefname{table}{Tab.}{Tabs.}
\Crefname{table}{Table}{Tables}
\newcommand{\model}{Apriel Voice 27B Preview}
\newcommand{\modelshort}{Apriel Voice}

\newcommand{\gemmafour}{Gemma4-12B-IT}
\newcommand{\ultravoxqwen}{UltraVox-v0.6 (Qwen)}
\newcommand{\audex}{Audex-30B-A3B}

\newcommand{\fluxgpt}{Flux + GPT-5.6}
\newcommand{\fluxgemma}{Flux + Gemma4-26B}
\newcommand{\parakeetgemma}{Parakeet + Gemma4-26B}
\newcommand{\parakeetgpt}{Parakeet + GPT-5.6}

\newcommand{\geminigeneral}{Gemini 3.8 Family}
\newcommand{\geminiss}{Gemini 3.8 Live Extended Thinking}
\newcommand{\gptrealtime}{GPT-realtime-2.1}

\newcommand{\tauvoice}{$\tau$-Voice}

\newcommand{\stageOneAudioHours}{32.2k}
\newcommand{\stageOneEpochs}{2}
\newcommand{\stageOneAudioExposure}{64.4k}
\newcommand{\stageTwoAudioHours}{61.5k}
\newcommand{\stageTwoEpochs}{3}
\newcommand{\stageTwoAudioExposure}{184.6k}

\newcommand{\wideappendixheader}[1]{#1}
\title{\model: An Open Model for Speech, Reasoning, and Tool Use}
\affiliation{ServiceNow}
\contribution{Contributors are listed in the \nameref{sec:contributors} section.}
\snowdata[Model]{\href{https://huggingface.co/ServiceNow-AI/Apriel-Voice-27B-preview}{Apriel Voice 27B Preview}}
\snowdata[Demo]{\href{https://servicenow-ai-apriel.hf.space/home}{Try Apriel Voice}}
\correspondence{\href{mailto:aman.tiwari@servicenow.com}{Aman Tiwari}\enspace\textperiodcentered\enspace\href{mailto:akshay.kalkunte@servicenow.com}{Akshay Kalkunte}}

\abstract{
A capable voice assistant must do more than recognize words: it must understand a request, reason over context, sustain a conversation, and act through tools. We present \model, an open speech-language model that unifies these abilities into a shared reasoning backbone while retaining text and image input modalities. Built on the post-trained Gemma 4 26B-A4B checkpoint, Apriel Voice processes speech directly and generates both responses and tool calls without requiring a separate dialogue model. A two-stage training procedure adapts the model to speech. Stage 1 establishes the speech interface. Stage 2 jointly adapts the projector and backbone using supervised fine-tuning, text replay, and source-dependent cross-entropy and forward Kullback--Leibler divergence supervision. Automatically constructed speech-conditioned dialogues provide conversational training data. Local teacher serving, overlapping teacher and student computation, and fused full-vocabulary losses enable effective online distillation. In thinking mode, Apriel Voice scores 98.90 on Big Bench Audio, 81.13 on Speech-BFCL, and 65.51\% task success on $\tau$-Voice, with a mean latency of 1.91 seconds. English EVA accuracy and experience pass@1 reach 51.27\% and 74.13\%. We serve \model{} through a custom platform to enable interactive voice interactions.

}

\begin{document}

\maketitle

\begin{figure*}[htbp]
  \centering
  \includegraphics[width=\textwidth]{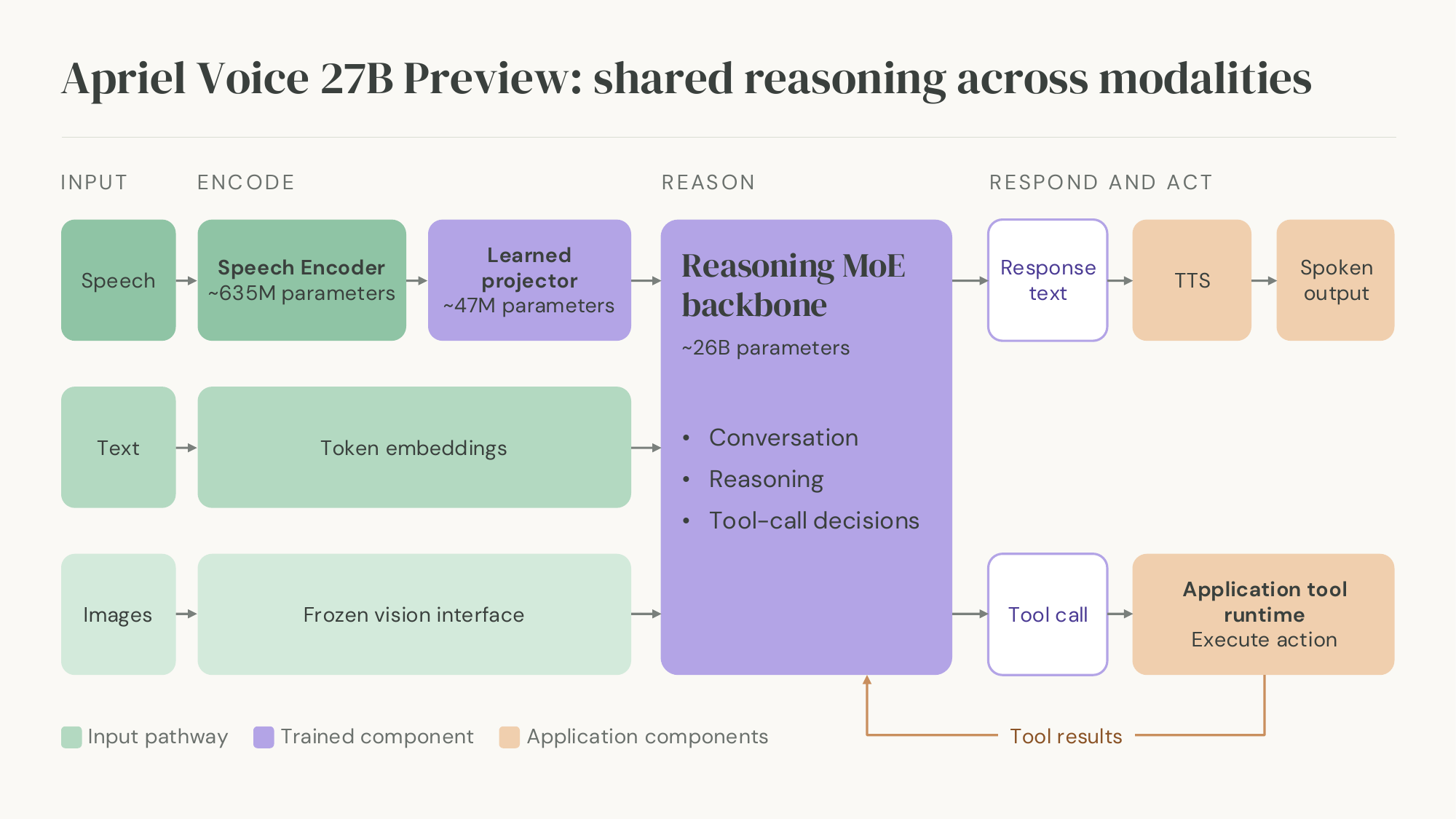}
  \caption{\textbf{One reasoning backbone across speech, text, and images.} Apriel Voice receives speech through an encoder and projector, retains the original text and image interfaces, and generates both responses and tool calls. Application components execute tools, return their results to the conversation, and stream response text to a replaceable synthesizer. The frozen teacher is used only during training.}
  \label{fig:architecture}
\end{figure*}
\clearpage

\section{Introduction}
An employee reporting a locked account may describe the problem aloud, share a screenshot, correct an identifier, and ask an assistant to open a ticket. The useful outcome is a resolved request: the assistant must connect the evidence, follow the conversation, choose an appropriate action, and interpret the result. Supporting such interactions requires speech understanding alongside the reasoning and tool use of a large language model (LLM).

Recognition accuracy is necessary but insufficient for voice agent success. A model may transcribe a request correctly yet lack the ability to follow constraints or invoke the necessary tools to complete the associated task. Likewise, strong text performance does not establish that the same model can reason reliably when the request arrives as audio. We therefore evaluate speech as an interface to general capability, covering reasoning, instruction following, conversation, and task completion alongside recognition and translation~\citep{mmaupro,eva,tauvoice}.

Voice systems connect these capabilities in several ways. A conventional \textbf{3-stage} cascade combines automatic speech recognition (ASR), an LLM, and text-to-speech (TTS) synthesis, passing a transcript to the reasoner. Speech-conditioned models instead expose the reasoner to learned acoustic features~\citep{slamasr,blsp,salmonn,ultravox}; native speech-generating models also integrate the output audio pathway~\citep{moshi,qwen3omni}. Another deployment pattern separates the spoken conversation from a backend model responsible for reasoning and tools~\citep{voicearchitectures}. These designs allocate speech understanding, reasoning, synthesis, and interaction control to different components.

We present \textit{\model}, or \textit{\modelshort}~in short, an open speech-language model built on the post-trained Gemma 4 26B-A4B checkpoint~\citep{gemma4,gemma4card}. \modelshort~uses the same reasoning backbone to understand speech, respond in conversation, and decide when and how to call tools. Text, images, and returned tool results remain part of its context (\cref{fig:architecture}). Speech synthesis is independently replaceable, allowing voices and deployment choices to change without retraining the reasoner.
We release \href{https://servicenow-ai-apriel.hf.space/home}{Apriel Live}, an interactive interface enabling practitioners to experience \modelshort~in real conversations.

The training problem is to extend an already capable decoder with a speech interface while preserving the behavior that makes it useful. We address it in two stages. Multilingual transcription, translation, and continuation first train a speech projector with the backbone frozen. Broad supervised fine-tuning (SFT) then updates the projector and backbone using speech tasks, internally generated conversations, tool interactions, and text replay. Source-dependent cross-entropy (CE) and forward Kullback--Leibler divergence (FKL) combine task targets with online supervision from a frozen copy of the original backbone. The teacher supports learning during adaptation and is absent at inference.

\paragraph{Contributions.}
This report brings together four contributions:
\begin{enumerate}
  \item \textbf{Broad speech-conditioned agentic capability.} A shared model handles spoken reasoning, instruction following, conversation, and tool-call generation while retaining text and image processing capabilities.
  \item \textbf{A staged adaptation recipe.} Speech interface learning with automatically constructed speech-conditioned dialogues and source-dependent CE/FKL supervision connects task acquisition with capability retention.
  \item \textbf{An efficient online-supervision implementation.} Local teacher serving, overlap with student computation, fused full-vocabulary losses, and coordinated checkpointing support the combined training workload.
  \item \textbf{Evaluation and interactive deployment.} Detailed results span speech, text, images, multilingual translation, service interactions, and enterprise workflows; Apriel Live demonstrates the corresponding application interface.
\end{enumerate}

In thinking mode, Apriel Voice achieves 98.90 on Big Bench Audio (BBA), 81.13 on BFCL, and 65.51\% mean task success on \tauvoice. English EVA accuracy and experience pass@1 reach 51.27\% and 74.13\%. Together, these results show a strong open-source (OSS) model whose speech capability extends to reasoning and action. The broader evaluation examines language coverage, reasoning-mode effects, and text and image retention. \Cref{tab:speechcomparison} summarizes the speech benchmarks; the following sections describe the method and detailed results.

\section{Related Work}
\subsection{Connecting audio perception to language reasoning}
Speech encoders provide transferable representations for recognition and downstream tasks~\citep{whisper}. SLAM-ASR studies a simple learned connection between a speech encoder and a language model~\citep{slamasr}; Ultravox develops a speech-conditioned architecture with frame stacking and a learned projector~\citep{ultravox}. SALMONN broadens the input scope to speech, audio events, and music~\citep{salmonn}. \model{}~follows this established encoder--projector approach. Its emphasis is the capability of the adapted model across reasoning, instruction following, and complete interactions, evaluated together with recognition.

\subsection{Transferring behavior across modalities}
BLSP trains a speech adapter using continuations generated from text prefixes, and BLSP-KD transfers next-token distributions between text and speech with token alignment and partial parameter adaptation~\citep{blsp,blspkd}. These methods are direct precedents for teaching a language model to respond to speech through behavioral supervision. \modelshort~builds on that principle with staged interface learning, broad second-stage SFT, selected speech distillation, and replay for a post-trained multimodal backbone. Supervision aligns the teacher's and student's response positions; it does not require acoustic frames to match transcript tokens one to one.

Our objective combines established forms of supervision: sequence targets~\citep{seqdistillation}, predictive-distribution distillation~\citep{distillation}, and preservation of earlier behavior during adaptation~\citep{lwf}. The contribution lies in how these signals are selected and implemented across the speech and retention mixture. Similarly, synthetic instruction generation has precedent in Self-Instruct~\citep{selfinstruct}. Our data pipeline converts internally generated task dialogues into speech-conditioned examples through automatic span selection, a consistent reference voice within each conversation, and conversion-fidelity checks.

\subsection{Speech generation and delegated voice systems}
LLaMA-Omni combines speech perception, an instruction-tuned language model, and a streaming speech decoder~\citep{llamaomni}. Moshi models concurrent user and assistant audio for full-duplex interaction~\citep{moshi}, while Qwen3-Omni integrates multimodal perception and speech generation in its Thinker--Talker design~\citep{qwen3omni}. These output-generation architectures differ from application-level delegation, in which a conversational voice model hands reasoning or tool work to a separate backend~\citep{voicearchitectures}.

\modelshort~places conversation, reasoning, and tool-call decisions in the same speech-conditioned backbone. Application components execute tools, return results to the same backbone, and synthesize response text. Turn detection and interruption control operate around the model.

\subsection{Evaluating and training capable voice agents}
EVA-Bench (EVA) measures both service accuracy and interaction experience~\citep{eva}, and $\tau$-Voice evaluates task completion under realistic audio conditions~\citep{tauvoice,tauvoiceblog}. $\tau^2$-Bench and EnterpriseOps-Gym extend the evaluation to text-agent interaction and stateful enterprise workflows~\citep{tau2,enterpriseops}. These benchmarks complement isolated recognition and reasoning tests by asking whether an agent reaches the required outcome.

For training, Fully Sharded Data Parallel (FSDP) distributes model state across workers~\citep{fsdp}, and fused-kernel systems such as Liger reduce the overhead of language-model operations~\citep{liger}. \modelshort~combines student sharding with local online-teacher execution, a lookahead schedule, and full-vocabulary CE/FKL kernels. We describe these mechanisms together with the scale and duration of the completed training run.

\section{Model Architecture}
\label{sec:modelarchitecture}
\subsection{A shared backbone for perception, reasoning, and action}
\model{} starts from the instruction-tuned Gemma 4 26B-A4B checkpoint, a mixture-of-experts model with approximately 26 billion total parameters and approximately four billion active parameters per token~\citep{gemma4,gemma4card}. This backbone accepts text and images. We add a continuous speech interface while retaining its output vocabulary, so spoken requests can elicit the same forms of response, reasoning, and tool calls as text requests.

Audio features, text tokens, image representations, and tool results condition the same backbone. The model generates either a response to the user or a tool call. The application executes tools, returns their results to the conversation, and sends response text to a separate speech synthesizer.

\begin{table}[htbp]
  \centering\small
  \begin{tabular*}{\linewidth}{@{\extracolsep{\fill}}lcc@{}}
    \toprule
    Component & Stage one & Stage two \\
    \midrule
    Speech encoder & Frozen & Frozen \\
    Speech projector & Trainable & Trainable \\
    Language backbone and router & Frozen & Trainable \\
    Original vision tower and bridge & Frozen & Frozen \\
    Online teacher & Frozen & Frozen \\
    Speech synthesizer & External & External \\
    \bottomrule
  \end{tabular*}
  \caption{\textbf{Component updates during speech adaptation.} The projector has approximately 47.7\,M parameters. The teacher supplies training supervision; the independently selected synthesizer is used at serving time and is not trained by this recipe.}
  \label{tab:modelcomponents}
\end{table}

\subsection{From acoustic frames to decoder embeddings}
A frozen Whisper large-v3-turbo encoder $E$ maps waveform $a$ to 1280-dimensional acoustic features~\citep{whisper,whisperturbo}. A learned projector maps these features into the decoder's embedding space. Inference uses the acoustic representations directly, without first producing a transcript. The projector follows the frame-stacking and multilayer-perceptron construction used by Ultravox~\citep{ultravox}.

Let $T_a$ be the number of encoder frames. For $E(a)\in\mathbb{R}^{T_a\times1280}$, stacking eight adjacent frames gives
\begin{equation}
 U=\operatorname{stack}_8(E(a))
 \in\mathbb{R}^{\lceil T_a/8\rceil\times10240}.
 \label{eq:stacking}
\end{equation}
The final incomplete group is padded. Stacking shortens the acoustic sequence presented to the decoder while retaining each group's frame features. A bias-free linear layer maps 10240 dimensions to 4096; a SwiGLU gate reduces this to 2048; a second bias-free layer projects into the 2816-dimensional decoder space:
\begin{align}
 Z &= \operatorname{SwiGLU}\!\left(\operatorname{RMSNorm}(U)W_1^\top\right),\\
 h_a &= \operatorname{RMSNorm}(Z)W_2^\top
       =P_\phi(E(a)).
 \label{eq:interface}
\end{align}
RMSNorm denotes root mean square normalization. SwiGLU splits its input into two equal parts and gates one with the swish activation of the other. The matrices contribute $10240\times4096+2048\times2816$ parameters. Including both normalization scale vectors gives exactly 47,722,496 projector parameters.

Projected features replace reserved audio positions in the input embedding sequence. Segment offsets and lengths locate each replacement, allowing multiple audio spans to be interleaved with text. Images enter through the original visual interface. Let $\theta$ and $\phi$ denote the backbone and projector parameters, respectively. With $c$ denoting the remaining conversation and multimodal context, the response distribution is
\begin{equation}
 p_{\theta,\phi}(y\mid a,c)
 =\prod_{t=1}^{|y|}p_\theta(y_t\mid y_{<t},h_a,c).
 \label{eq:generation}
\end{equation}
The existing output vocabulary supports the same response and tool-call formats for speech and text inputs.

\subsection{Adapting the decoder while preserving its interfaces}
Stage one trains the speech projector with the encoder and backbone fixed. Stage two updates the projector, language backbone, and mixture-of-experts router together (\cref{tab:modelcomponents}). A frozen copy of the starting checkpoint supplies online teacher distributions during training and is removed from the inference graph.

Text examples continue to use the original token embeddings and output vocabulary. On text-only batches, a zero-valued dependency keeps the projector in the distributed computation graph without adding speech information. The original vision tower and bridge remain frozen in both stages. Their outputs nevertheless pass through a decoder that changes during stage two, so visual retention must be measured empirically. The text and image evaluations in \cref{sec:results} measure performance after adaptation.

\section{Training Recipe}
\label{sec:trainingrecipe}
\begin{figure*}[htbp]
  \centering
    \includegraphics[width=\textwidth]{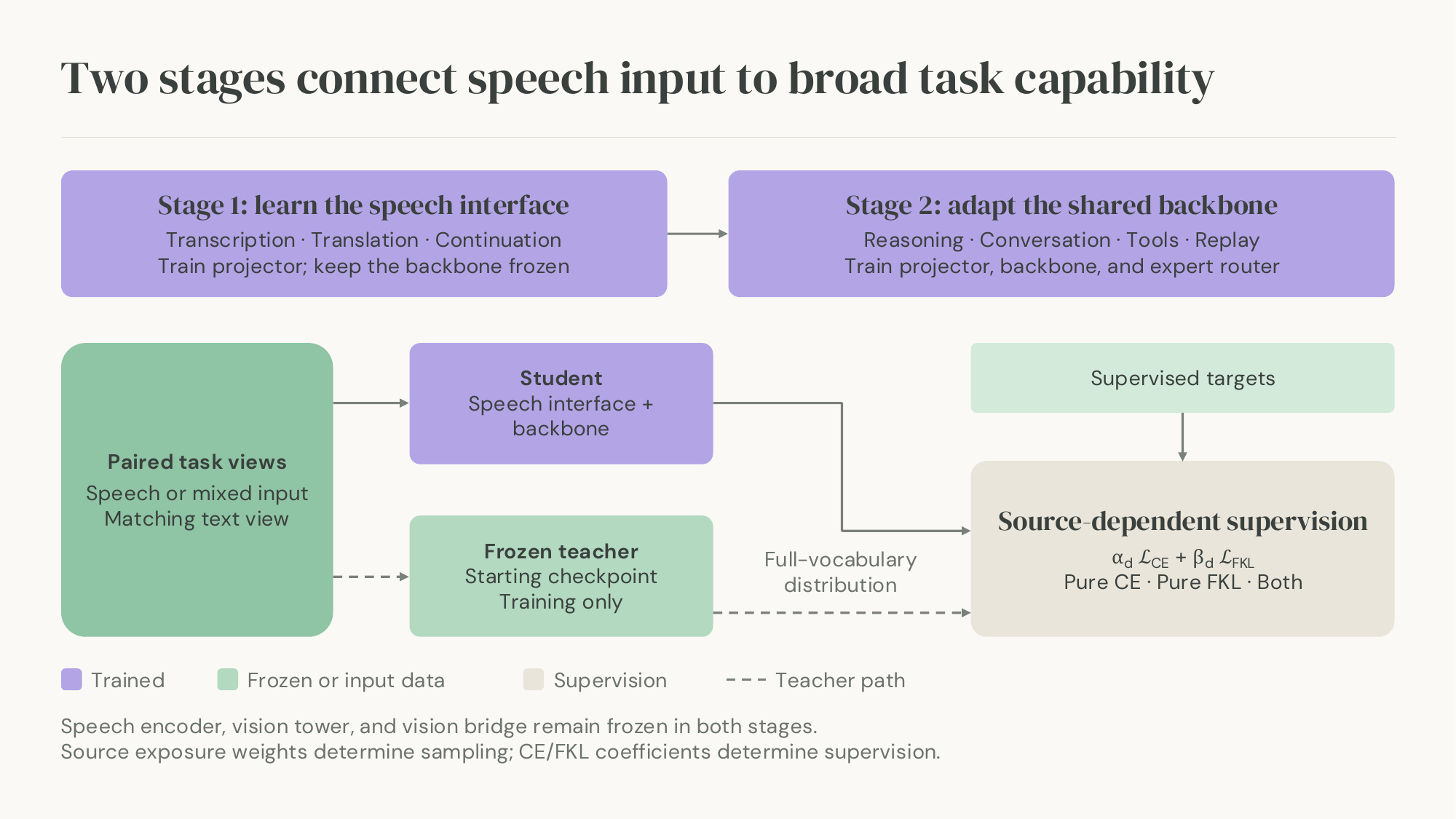}
  \caption{\textbf{Learning a speech interface, then broadening task capability.} Stage one trains the projector with the backbone fixed. Stage two updates both on speech tasks, synthetic conversations, tool interactions, and text replay. Source-dependent CE, FKL, or combined supervision supplies task targets and teacher distributions. The speech encoder and original visual components remain frozen.}
  \label{fig:training}
\end{figure*}

\subsection{Two stages with complementary goals}
Speech adaptation must solve two related problems: make acoustic input intelligible to an existing decoder, and teach that decoder to use speech in complete tasks. We separate them in the training recipe (\cref{fig:training}). Stage one learns the input interface through multilingual transcription, translation, and continuation. Stage two broadens the task distribution and updates the backbone, with explicit supervision for acquiring new behavior and retaining earlier capabilities.

Stage one uses 18.17\,M training examples across 1,164 dataset configurations grouped under 152 names. These include task-specific subsets and variants of Common Voice~\citep{commonvoice}, GigaSpeech~\citep{gigaspeech}, People's Speech~\citep{peoplesspeech}, LibriSpeech~\citep{librispeech}, and CoVoST~2~\citep{covost2}. Transcription predicts spoken content, translation changes its language, and continuation trains response prediction beyond verbatim recognition.

Stage two combines mathematics, science, general conversation, speech tasks, noisy interactions, tool use, and text replay. Conversational data cover clarification, disambiguation, confirmation before updates, and speech surrounding tool execution. These examples extend training from isolated utterances to interactions in which the next response depends on prior context and actions.

\subsection{Stage one: learning the input interface}
We train the projector with the speech encoder, language backbone, vision tower, and vision bridge frozen. Teacher supervision transfers the starting decoder's predictive behavior to speech-conditioned contexts. The projector learns to map speech into representations the fixed decoder can use.

A bin is a packed group of training examples. Two completed epochs each cover \stageOneAudioHours~hrs of audio. The mixture contains 405,931 raw bins and 476,747 effective bins after exposure weighting, with 99.2\% token fill. Per epoch, it contributes approximately 1.9367\,B total tokens and 631.7583\,M loss-bearing tokens. Packing accommodates many short utterances efficiently; only the designated prediction positions carry direct supervision.

\subsection{Stage two: broad task learning with retention}
Stage two updates the projector and language backbone, including the mixture-of-experts router. The speech encoder and original visual components remain frozen. Three epochs each contain \stageTwoAudioHours~hrs of audio plus text data.

Each source selects CE, FKL, or a weighted combination. CE teaches target responses, reasoning, and tool behavior. FKL transfers teacher behavior on selected speech tasks and constrains changes on retention replay. The two signals can therefore serve both modality adaptation and capability preservation; FKL is not restricted to text-only retention. \Cref{sec:objectives} formalizes their interaction.

\subsection{Packing and exposure accounting}
Each packed bin stores position resets, target masks, audio offsets, and teacher/student segment boundaries. These fields identify corresponding response regions even when the speech and text contexts have different lengths. Source exposure weights determine how often an example is encountered; loss coefficients determine the supervision applied when it is encountered.

For source $d$ with raw-bin count $B_d$ and exposure multiplier $w_d$, its normalized effective-bin contribution is
\begin{equation}
 \pi_d=\frac{w_d B_d}{\sum_{d'}w_{d'}B_{d'}}.
 \label{eq:exposure}
\end{equation}
Row counts, raw-bin counts, and weighted exposure consequently describe different aspects of the mixture. The stage-one statistic of 44.8 samples per bin uses raw bins; effective bins include exposure weighting.

\begin{table}[htbp]
  \centering\small
  \begin{tabular*}{\linewidth}{@{\extracolsep{\fill}}lr@{}}
    \toprule
    Training quantity & Value \\
    \midrule
    Total rows & 17.56\,M \\
    Packed bins & 1.56\,M \\
    Average token-fill efficiency & 92.5\% \\
    Tokens / loss-bearing tokens per epoch & 17.5\,B / 2.6\,B \\
    Completed epochs / total token exposure & 3 / 52.6\,B \\
    Nodes / teacher GPUs / student GPUs & 8 / 16 / 48 \\
    Wall-clock duration & $\approx$3 days \\
    \bottomrule
  \end{tabular*}
  \caption{\textbf{Stage-two training scale.} Rows denote training examples; bins include exposure weighting. M/B: million/billion. The duration describes the optimized training run.}
  \label{tab:training}
\end{table}

Stage two provides $17.5491\,\mathrm{B}\times3=52.6473\,\mathrm{B}$ total tokens of training exposure, including approximately 7.88574\,B loss-bearing tokens (\cref{tab:training}). Audio is counted separately: stage one repeats 32,213 hours twice, and stage two repeats 61,540 hours three times (\cref{tab:audiohours}). Text contributes tokens without an assigned audio duration. These totals measure repeated training exposure, not unique data volume.

\section{Data Construction}

\subsection{Constructing speech-conditioned conversations}
\label{sec:dataconstruction}
Our custom task and conversation data are generated synthetically in-house. We first construct text interactions containing user requests, assistant responses, and, where relevant, tool calls and returned results. This representation makes the intended task, role structure, and target behavior explicit. An automated conversion pipeline then renders selected user content as audio through reference-conditioned TTS and voice cloning.

Spoken requests can refer to material best retained as text, such as a code fragment, table, formula, or case-sensitive identifier. The conversion therefore supports both whole-turn speech and mixed text/audio turns. System instructions, assistant targets, and tool messages remain text, preserving the supervision for conversation and action.

\subsection{Automatic span selection and consistent voices}
The selector supports five strategies: speak the complete user turn; speak a question while keeping its context in text; speak a generative instruction; speak content requested for reading or recitation; or speak several separated spans. Its prompt favors retaining structured and case-sensitive material as text. A configurable override adds whole-turn conversions.

For selected source intervals $I_1,\ldots,I_K$ in user turn $u$, and a reference voice $r_D$ assigned to conversation $D$, conversion takes the form
\begin{align}
 a_j &= \operatorname{TTS}(N(u[I_j]),r_D),\\
 \widetilde u &= \operatorname{Interleave}
   \bigl(u,\{(I_j,a_j)\}_{j=1}^{K}\bigr),
 \label{eq:dataconstruction}
\end{align}
where $N$ converts written expressions into speakable form. Each selected span must match a source substring before it can be replaced. Unselected text retains its original position. This gives the student an acoustic task view and the teacher a corresponding text view.

\begin{figure*}[htbp]
  \centering
    \includegraphics[width=\textwidth]{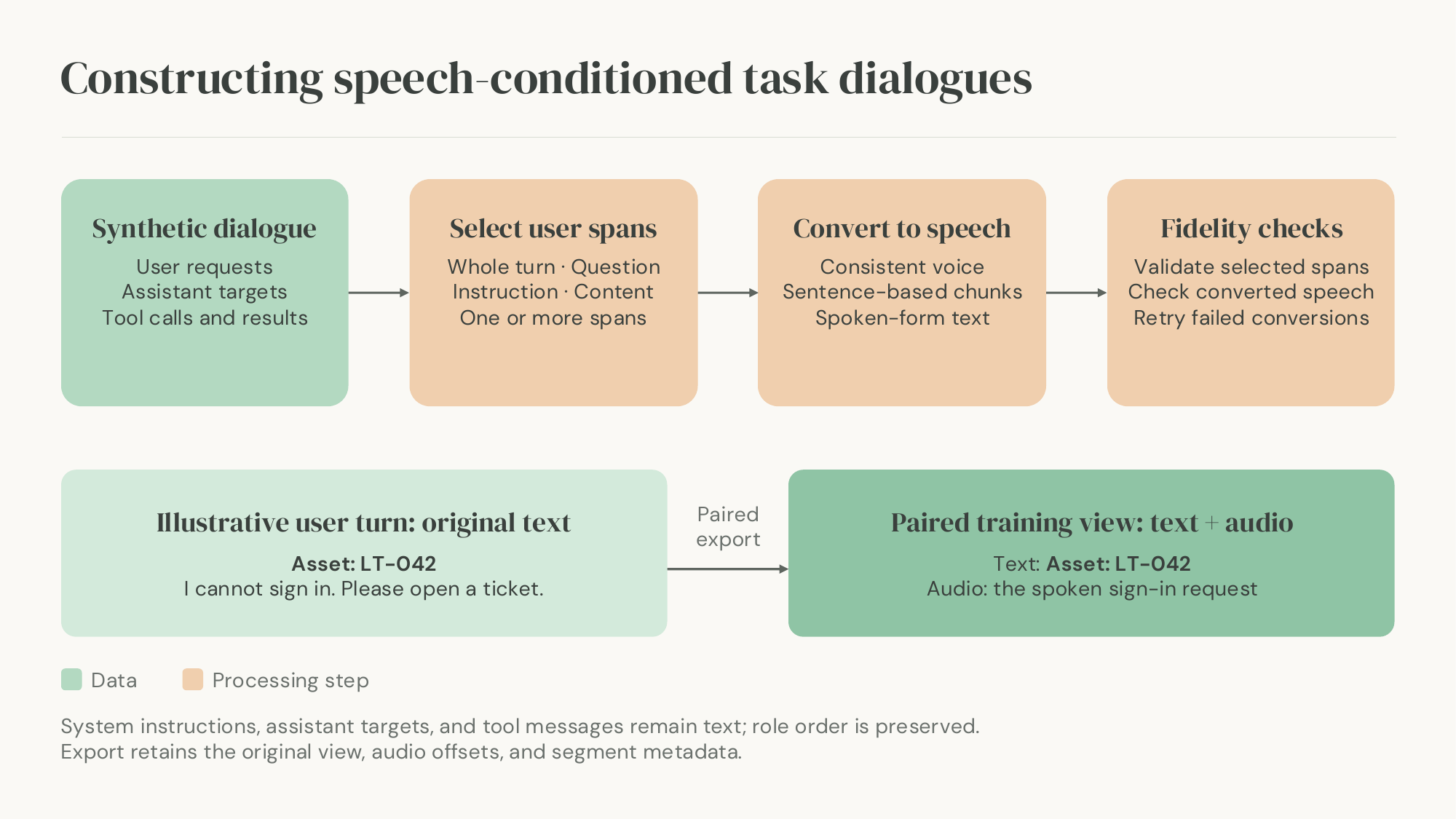}
  \caption{\textbf{From synthetic task dialogues to speech-conditioned training examples.} The pipeline selects user spans, normalizes their spoken form, synthesizes them with a conversation-level reference voice, and applies fidelity checks. Assistant targets and tool messages remain text. The illustrated asset-management dialogue is schematic.}
  \label{fig:datageneration}
\end{figure*}

The same reference utterance conditions all synthesized turns and chunks within a conversation. Long spans are split into sentence-oriented chunks subject to duration checks. Contiguous chunks can be joined, while separated spans retain their position among the text. Speaker-aware scheduling and resumable processing preserve the voice assignment during generation.

\subsection{Conversion fidelity and paired export}
The pipeline supports three fidelity checks: validation of source spans before conversion, spoken-form judging after normalization, and recognition-based judging after synthesis. Failed conversions and chunks can be retried within a configured budget, and an optional turn-level faithfulness check provides further validation. These checks reduce mismatches between the intended input and synthesized speech; they do not establish the correctness of a synthetic task or its target answer.

Export retains role order, assistant supervision, tool context, audio offsets, and the original text view. Offline task generation and online distillation provide different signals: synthetic responses can supply CE targets without FKL, while selected examples receive teacher distributions during optimization. \Cref{sec:datagenappendix} gives further details of selection, chunking, and export.

\section{Task Learning and Capability Retention}
\label{sec:objectives}
\subsection{Learning from targets and predictive distributions}
Task targets specify what the model should generate. Teacher distributions express the starting decoder's relative preferences among possible next tokens. Combining these signals allows the model to learn new behavior while retaining selected predictions of the post-trained model~\citep{seqdistillation,distillation,lwf}.

Let $(x_i,y_i)$ be a student example from source $d$, with teacher-compatible context $x_i^T$ when distillation is enabled. The student can receive audio, text, and images; for speech tasks, the teacher receives the corresponding text view. Both predict the supervised response from its target prefix. Their input lengths can differ, because distribution matching concerns response positions rather than a one-to-one alignment of acoustic frames and transcript tokens.

The source-dependent objective is
\begin{equation}
 \mathcal{L}_i=
 \alpha_d\,\mathcal{L}_{\mathrm{CE},i}
 +\beta_d\,\mathcal{L}_{\mathrm{FKL},i},
 \label{eq:loss}
\end{equation}
where $\alpha_d,\beta_d\geq0$ and at least one coefficient is positive. For a common nonempty set of supervised positions $S_i$, the losses are
\begin{align}
 \mathcal{L}_{\mathrm{CE},i}
 &=-\frac{1}{|S_i|}\sum_{t\in S_i}\log p_{\theta,t}(y_{i,t}),\\
 \mathcal{L}_{\mathrm{FKL},i}
 &=\frac{1}{|S_i|}\sum_{t\in S_i}\sum_{v\in\mathcal{V}}
 p_{0,t}(v)\log\frac{p_{0,t}(v)}{p_{\theta,t}(v)}.
 \label{eq:fkl}
\end{align}
Here $p_0$ is the frozen Gemma teacher and $\mathcal{V}$ is the shared output vocabulary. The speech projector influences $p_\theta$ through the input embeddings. Stage-two full-vocabulary FKL uses temperature one.

\subsection{Choosing the supervision for each source}
Pure CE uses $\beta_d=0$; pure FKL uses $\alpha_d=0$; combined supervision uses both positive coefficients. Their sum need not be one. \Cref{tab:lossweights} summarizes these routes. A source's loss settings are distinct from the exposure weights that determine its sampling frequency.

This flexibility serves two purposes. Task targets teach desired answers, reasoning, and tool behavior, including targets constructed offline. Teacher distributions transfer existing behavior to speech and preserve it on replay contexts. A source can use either signal independently or combine them.

\subsection{Interpreting the combined objective}
On a token supervised by both losses with the same mask and normalization, let $e_{y_t}$ denote the one-hot target. The combined objective is, up to a teacher-dependent constant, a scaled cross-entropy with the soft target
\begin{equation}
 q_{d,t}=(1-\lambda_d)e_{y_t}+\lambda_d p_{0,t},\qquad
 \lambda_d=\frac{\beta_d}{\alpha_d+\beta_d}.
 \label{eq:softtarget}
\end{equation}
Here $H(q,p)=-\sum_v q(v)\log p(v)$ denotes cross-entropy and $H(p)=H(p,p)$ denotes entropy. The exact token loss is $(\alpha_d+\beta_d)H(q_{d,t},p_{\theta,t})-\beta_dH(p_{0,t})$. Thus $\lambda_d$ sets the target mixture, while $\alpha_d+\beta_d$ scales its contribution. With $z_{\theta,t}$ denoting the student logits, the corresponding gradient is
\begin{equation}
 \nabla_{z_{\theta,t}}\mathcal{L}_i=
 \frac{\alpha_d(p_{\theta,t}-e_{y_t})
       +\beta_d(p_{\theta,t}-p_{0,t})}{|S_i|}.
 \label{eq:objectivegradient}
\end{equation}
Task supervision favors the target response; teacher supervision penalizes divergence from the teacher on the same context.

The derivation assumes common supervised positions and normalization. In packed execution, response extraction, causal shifts, masking, and reduction across segments determine the effective token weights. If CE and FKL use different masks or normalizers, the mixture interpretation applies where both losses provide supervision, with mixture weights determined by their effective token weights.

The objective constrains behavior on the contexts encountered during training. Retention on other inputs therefore remains an empirical question, especially when the teacher is incorrect or poorly suited to a new task. \Cref{sec:results,sec:discussion} examine the resulting balance between task learning and capability retention.

\section{Efficient Training with an Online Teacher}
\label{sec:trainingsystems}
Online full-vocabulary distillation adds a teacher forward pass and distribution transfer to the student's training workload. The implementation distributes teacher inference across local workers, overlaps it with student computation, and reduces temporary storage for loss evaluation.

\subsection{Local teacher workers and sharded students}
Teacher inference and student training have different resource requirements. The teacher performs forward computation; the student also stores optimizer state and executes backward computation. We assign them separate workers and use FSDP to shard student state~\citep{fsdp}.

Student workers send teacher contexts and position information to assigned teacher workers through GPU point-to-point communication. The full-vocabulary path returns teacher logits in bfloat16 (BF16) format. This traffic stays within a node, while student workers use distributed collectives across their own process group. The same local allocation can therefore be repeated across nodes.

Stage-two training uses eight nodes, each with two teacher GPUs and six student GPUs: 16 teacher and 48 student NVIDIA H100 80 GB HBM3 GPUs overall. Training completed in approximately three days (\cref{tab:training}).

\subsection{Overlapping teacher and student computation}
\begin{figure*}[htbp]
  \centering
  \includegraphics[width=\textwidth]{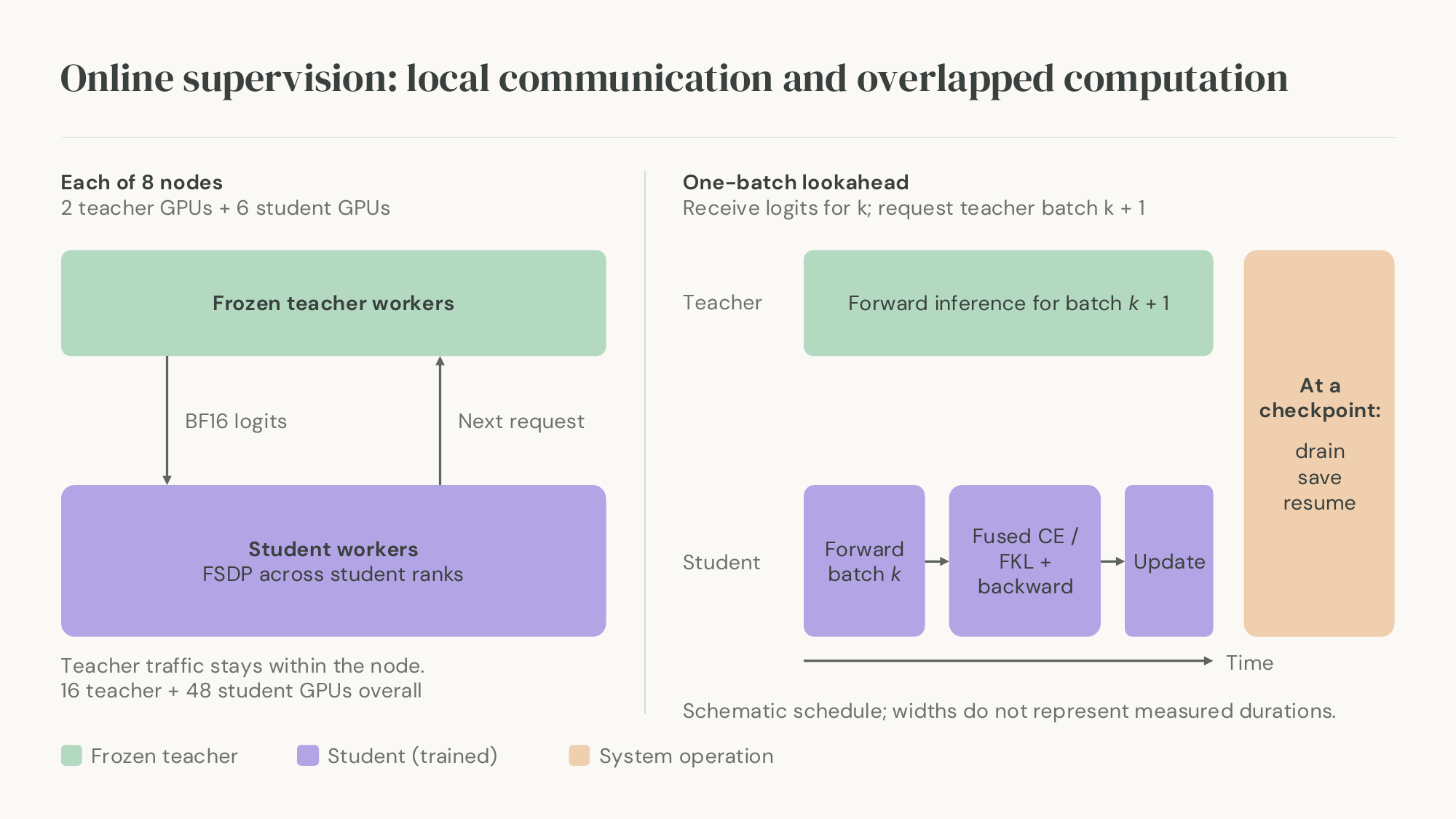}
  \caption{\textbf{Local online supervision with one-batch lookahead.} Schematic schedule in which teacher inference for the next batch overlaps the student's current training work. Teacher--student communication stays within each node; student state is sharded across training workers. Workers complete pending teacher requests before saving checkpoints.}
  \label{fig:teacherserving}
\end{figure*}

The training loop uses one-batch lookahead. After receiving the current teacher result, it issues the next request before executing the current student computation. Teacher inference for batch $k+1$ can then overlap student forward and backward computation for batch $k$ (\cref{fig:teacherserving}). The first teacher request primes the pipeline.

An idealized steady-state lower bound is
\begin{equation}
 T_{\mathrm{step}}\geq
 \max\{T_{\mathrm{student}},T_{\mathrm{teacher,amortized}}\},
 \label{eq:overlapbound}
\end{equation}
excluding transfer, synchronization, and data-loading overhead. The amortized teacher term accounts for one worker serving multiple students. Speech-length variation and queueing determine how much of the potential overlap is realized.

Checkpointing must preserve this communication schedule. Students synchronize the checkpoint decision, drain outstanding teacher requests, save state, and restart the lookahead pipeline. Coordinating these transitions prevents workers from entering incompatible communication or checkpoint phases.

\subsection{Fused CE and full-vocabulary forward KL}
For $B$ sequences, $L$ response positions per sequence, and vocabulary size $V$, a dense logit tensor contains $BLV$ entries. Full-vocabulary matching requires teacher and student logits, but naively retaining additional probabilities and log-probabilities in 32-bit floating-point (FP32) format further expands the working set. We use fused reductions and vocabulary tiling to bound these intermediates, following the broader motivation of fused language-model kernels~\citep{liger}.

For the teacher and student logits separately, the FKL kernels first compute a normalizer over the complete vocabulary,
\begin{equation}
 \log Z_t=\log\sum_{v\in\mathcal{V}}\exp z_{t,v},
 \label{eq:lognormalizer}
\end{equation}
then reuse it while reconstructing probabilities in tiles. Every tile uses that same row normalizer, preserving the full-vocabulary distribution. The backward pass evaluates the teacher-matching term $p_{\theta,t}-p_{0,t}$ from \cref{eq:objectivegradient}; the frozen teacher receives no gradient. A fused CE path similarly avoids retaining unnecessary probability intermediates.

The loss wrapper also chunks long response regions along the sequence dimension. Within a region, chunk losses are combined using their supervised-token counts, so a short final chunk receives proportionate weight. Full logits remain materialized: the memory-saving mechanism bounds temporary work and avoids dense FP32 probability intermediates, rather than eliminating logit storage.

\subsection{Optimization settings}
Stage two uses BF16 mixed precision, activation checkpointing, and fused multi-tensor AdamW. The peak learning rate is $4\times10^{-6}$, with warmup over 20\% of the schedule, cosine decay to 10\% of the peak, and gradient clipping at 1.0.

\Cref{fig:loss,fig:lossfull} show the combined objective and its CE/FKL components over training. The curves track optimization across the mixed task distribution.

\section{Evaluation Design}
\label{sec:evaluation}
\subsection{Questions and coverage}

We structure our evaluation around three major research questions to establish whether a unified speech-language backbone trained with our two-stage recipe can achieve competitive performance compared to leading open-source and proprietary systems across reasoning, end-to-end task completion, and multimodal capability retention.

\paragraph{RQ1: Whether \modelshort~supports audio reasoning, instruction following, and tool use beyond foundational speech capabilities.} One of the core contributions of speech-language models lies in their ability to process spoken instructions, reason over problems, and generate tool calls, not merely perform foundational speech recognition and translation. Therefore, we target evaluations across different foundational tasks (speech recognition, speech translation) and audio reasoning tasks, including instruction following (MT-Bench~\citep{mtbench}, IFEval~\citep{ifeval}), reasoning over spoken questions (Big Bench Audio~\citep{bigbenchaudio}, Speech-GSM8K~\cite{gsm8k}), and tool-call capability via BFCL~\citep{bfcl}. In addition, we continue to validate the foundational speech capabilities of the model: speech recognition (LibriSpeech~\citep{librispeech}, $\mu$-Bench~\citep{mubench}) and speech translation (CoVoST~2~\citep{covost2}). Accuracy and reasoning scores are reported on a 0--100 scale; WER can exceed 100\%.

\paragraph{RQ2: Whether those abilities translate into effective conversation and complete workflows.} Beyond single-turn reasoning and instruction following, production systems must sustain multi-turn conversations, maintain context, and execute complete tasks through speech. To this end, EVA measures service accuracy and conversational experience, while $\tau$-Voice evaluates task completion through speech~\citep{eva,tauvoice}. We complement these with $\tau^2$-Bench for text-based agent interaction evaluation, and EnterpriseOps-Gym for testing stateful tool workflows~\citep{tau2,enterpriseops}. Together, these benchmarks distinguish between isolated correct answers and real-world complete task execution.

\paragraph{RQ3: How text and image performance changes after adaptation.}
Text-input evaluations cover multidisciplinary knowledge and reasoning with MMLU-Pro~\citep{mmlupro}, graduate-level science with GPQA Diamond~\citep{gpqa}, and mathematical problem solving with the American Invitational Mathematics Examination (AIME), MATH-500~\citep{math}, and the American Mathematics Competitions (AMC). Paired text/audio evaluations also test instruction following and tool use. Image evaluations cover optical character recognition (OCR), visual mathematics, multidisciplinary reasoning, charts, instruction following, and hallucination-related errors. Internal Gemma evaluations use the starting Gemma 4 26B-A4B backbone; published model-card results provide additional comparisons under their own evaluation settings.

\subsection{Evaluation implementations}
We use AU-Harness for BFCL, speech-recognition error rates, and the applicable audio-to-text instruction, conversation, and translation evaluations~\citep{auharness}. It supplies task definitions, prompts, preprocessing, output parsing, and scoring. EVA, $\tau$-Voice, and Audio Reasoning test suite use their official benchmark implementations~\citep{eva,tauvoice,auharness}.

\subsection{Model Inference Setup}
\modelshort's default decoding settings are temperature 1.0, top-$k$ 64, and top-$p$ 0.95, subject to benchmark-specific generation logic. Recognition and translation evaluations---LibriSpeech, FLEURS, $\mu$-Bench, and CoVoST~2---use temperature 0.4 with the same top-$k$ and top-$p$. Thinking and non-thinking denote whether the checkpoint's reasoning mode is enabled; results are reported separately when both are available. \modelshort~uses thinking mode for EVA, $\tau$-Voice, and image/document evaluations.

EVA evaluates \modelshort~with Cartesia Sonic 3.5 text-to-speech~\citep{sonic35}; the $\tau$-Voice evaluation also uses Cartesia synthesis. These results describe the complete voice system. The internal $\tau^2$ comparison uses GPT-5.2 as the user simulator for both \modelshort~and Gemma. Public comparisons use the model variants, reasoning settings, synthesizers, task limits, and simulators documented in their cited evaluations.

\subsection{Metric Details}

\paragraph{Error rates and translation metrics (RQ1).}
Word error rate (WER) counts transcript errors relative to the reference length; lower is better. For substitutions $S$, deletions $D$, insertions $I$, and $N$ reference words,
\begin{equation}
 \operatorname{WER}=100\,\frac{S+D+I}{N}.
 \label{eq:errorrate}
\end{equation}
Insertions can produce WER values above 100\%. We evaluate English speech recognition using standard WER metrics on LibriSpeech, GigaSpeech, AMI, $\mu$-Bench, FLEURS, and Earnings21. WER is computed after text normalization via AU-Harness~\citep{auharness} for consistency purposes. Lower values indicate better performance for WER. For speech translation tasks, CoVoST~2 uses BLEU (bilingual evaluation understudy), which measures $n$-gram overlap with reference translations and penalizes excessively short outputs. BLEU is reported on a 0--100 scale; higher is better.

\paragraph{Accuracy, experience, and responsiveness (RQ2).}
Pass@1 is the fraction of evaluation cases that succeed in a single attempt. EVA-A requires speech fidelity, faithfulness to the interaction, and task completion to pass jointly; EVA-X requires concise responses, effective conversation progression, and appropriate turn-taking~\citep{eva}. These composite scores are not averages of the component scores. Component means describe performance on each criterion.

Turn-taking explicitly incorporates timing: ordinary response delays, overlap and recovery after interruptions, yielding when the user interrupts, and conversation completion. Tool-call turns use more lenient timing thresholds. EVA-X therefore measures conversational responsiveness as part of the complete user experience. For \modelshort, $\tau$-Voice latency is the delay until the spoken response is available after TTS, including response generation and speech synthesis. We report it in seconds alongside task success.

\begin{figure}[htbp]
  \centering
  \includegraphics[trim={1.2cm 3cm 2.3cm 1.5cm}, clip, width=\linewidth]{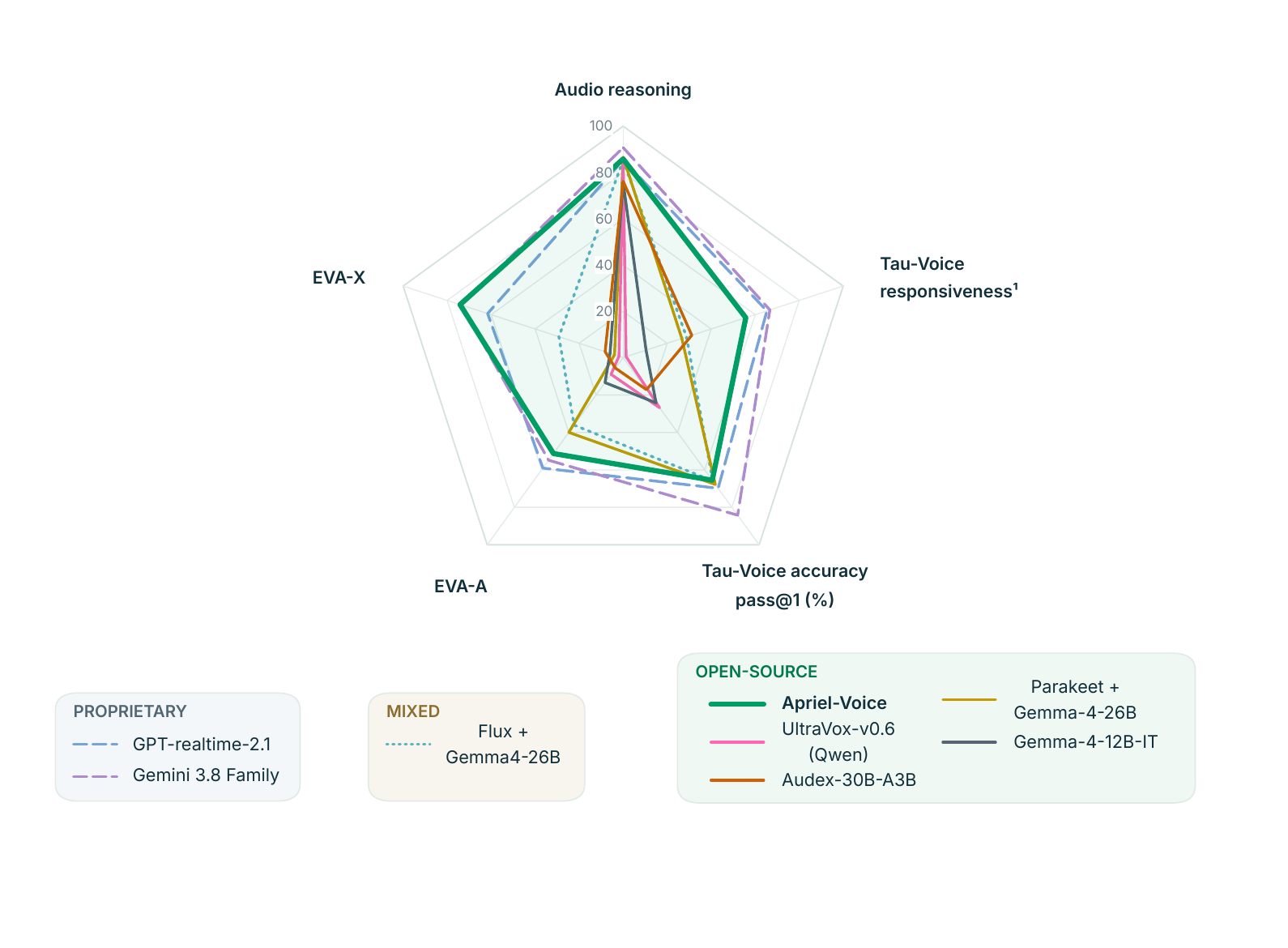}
  \caption{\textbf{Multi-dimensional comparison of voice model performance.}
The radar plot summarizes each model across five axes: audio reasoning, \tauvoice~responsiveness, \tauvoice~accuracy, EVA-A, and EVA-X. Higher values are better on all axes. \tauvoice~accuracy is measured as pass@1 percentage, while EVA-A and EVA-X report the model's accuracy and experience scores on the EVA benchmark. Tau-Voice responsiveness converts response latency into a bounded 0--100 score as described in Equation~\ref{eq:latency_score}, where lower latency receives a higher score. \textbf{\modelshort} is shown as the green reference profile, and the remaining profiles are grouped by source category.}
  \label{fig:radar_comparison}
\end{figure}

To enable fair comparison across models with varying latency characteristics, we normalize response latency into a 0--100 \textit{Responsive Score} as observed in Figure \ref{fig:radar_comparison}. For a given response time $t$ (in seconds), we compute:

\begin{equation}
R_{\text{score}}(t) = \operatorname{clip}\left(100 \cdot \frac{t_{\max} - t}{t_{\max} - t_{\min}}, 0, 100\right), \quad t_{\min} = 0.25\text{s}, \quad t_{\max} = 4.0\text{s}
\label{eq:latency_score}
\end{equation}

where $t_{\min} = 0.25\text{s}$ represents the excellent-latency threshold (score of 100) and $t_{\max} = 4.0\text{s}$ represents the poor-latency threshold (score of 0). Latencies between these bounds are linearly interpolated; the $\operatorname{clip}$ function ensures all scores remain in $[0, 100]$.

\paragraph{Domain aggregation and workflow completion (RQ2).}
For $D$ domains, let $s_d$ be the success rate and $n_d$ the number of evaluation cases in domain $d$. Equal-domain averaging and pooling over cases are
\begin{equation}
 s_{\mathrm{macro}}=\frac{1}{D}\sum_{d=1}^{D}s_d,
 \qquad
 s_{\mathrm{pool}}=\frac{\sum_{d=1}^{D}n_ds_d}{\sum_{d=1}^{D}n_d}.
 \label{eq:macro}
\end{equation}
EVA performance and the \modelshort~$\tau$-Voice summary are computed as equal-domain means over three domains. The public EVA leaderboard uses pooled pass@1.

EnterpriseOps-Gym distinguishes strict task completion from partial verifier success. For trajectory $j$ with binary checks $v_{j,k}$,
\begin{equation}
 q_j=\prod_{k=1}^{K_j}v_{j,k},\qquad
 r_j=\frac{1}{K_j}\sum_{k=1}^{K_j}v_{j,k}.
 \label{eq:entops}
\end{equation}
Strict completion requires $q_j=1$, so every check must pass; $r_j$ summarizes partial progress within a trajectory. Verifier-success rates aggregate check outcomes across evaluation tasks. Oracle mode supplies the tools relevant to the task; additional-tool conditions also supply distractors.

\begin{table*}[t]
\centering
\small
\resizebox{0.5\textwidth}{!}{
\begin{tabular}{lccc}
\toprule
\textbf{Model} & \textbf{Source} & \textbf{Type} & \textbf{Access} \\
\midrule
\geminigeneral & Proprietary & S2S & Closed \\
\gptrealtime & Proprietary & S2S & Closed \\
\fluxgpt & Proprietary & 3-stage & Closed \\
\fluxgemma & Mixed & 3-stage & Partial \\
\parakeetgpt & Mixed & 3-stage & Partial \\
\parakeetgemma & Open-source & 3-stage & Open \\
\ultravoxqwen & Open-source & 2-stage & Open \\
\gemmafour & Open-source & 2-stage & Open \\
\audex & Open-source & 2-stage & Open \\
\midrule
\modelshort & \textbf{Open-source} & \textbf{2-stage} & \textbf{Open} \\
\bottomrule
\end{tabular}
}
\caption{\textbf{Summary of baseline models: source, type, and access.} We compare \modelshort~against proprietary systems (S2S and 3-stage), mixed systems (3-stage cascaded), and open-source models (2-stage and 3-stage). 2-stage systems project speech into a language backbone; 3-stage systems cascade ASR, LLM, and TTS; S2S generates spoken output directly.}
\label{tab:baseline_models}
\end{table*}

\subsection{Baseline Comparison}
We deliberately evaluate \modelshort~against a representative set of speech-language models spanning distinct architectural paradigms: cascaded 3-stage pipelines (ASR-LLM-TTS), 2-stage speech-conditioned models, and end-to-end speech-to-speech (S2S) systems. This diversity provides a complete picture of the landscape \modelshort~operates within. \textit{Proprietary systems}, \textit{Gemini 3.8 Family} models, \textit{\gptrealtime}, strong ASR+LLM model combination \textit{\fluxgpt} establish strong performance references but remain closed. \textit{Mixed systems} like \textit{\fluxgemma}, \textit{\parakeetgpt} combine proprietary and open-source components, demonstrating hybrid approaches. \textit{Open-source models}---\textit{Parakeet + Gemma 4 26B}, \textit{\ultravoxqwen}, \textit{\gemmafour}, \textit{\audex}~\cite{audex}---represent the category \modelshort~targets, capturing recent progress in open speech-language capability. A summary of the baseline models is provided in Table \ref{tab:baseline_models}.

\section{Results and Analysis}
\label{sec:results}

Evaluating speech-language models for real-world deployment requires comprehensive assessment across multiple dimensions---from foundational speech understanding to complex reasoning and end-to-end task execution. Therefore, we structure our evaluation around three key dimensions: (1) reasoning and instruction-following beyond foundational speech tasks, (2) conversational capabilities and task completion, and (3) retention of text and image capabilities after speech adaptation. The essential multi-dimensional considerations are captured in Figure \ref{fig:radar_comparison}.

\begin{table}
\centering
\resizebox{\textwidth}{!}{%
\begin{tabular}{lccc>{\columncolor{blue!10}}c>{\columncolor{blue!10}}c>{\columncolor{blue!10}}c>{\columncolor{blue!10}}c>{\columncolor{green!20}}c}
\toprule
\textbf{Metric} & \multicolumn{3}{c}{\textbf{Proprietary}} & \multicolumn{5}{c}{\textbf{Open-Source}} \\
\cmidrule(lr){2-4} \cmidrule(lr){5-9}
& \textbf{Flux} & \textbf{\geminigeneral} & \textbf{\gptrealtime} & \textbf{Parakeet} & \textbf{\ultravoxqwen} & \textbf{\gemmafour} & \textbf{\audex} & \textbf{\modelshort} \\
\midrule
\multicolumn{9}{l}{\textit{Speech Recognition (WER \%, lower is better)}} \\
LS test-clean & 3.51 & 2.10 & 1.85 & 1.65 & 6.11 & 5.77 & 1.36 & 2.23 \\
LS test-other & 8.26 & 4.42 & 4.34 & 3.14 & 8.08 & 18.51 & 3.08 & 5.15 \\
GigaSpeech & 10.68 & 10.16 & 10.57 & 10.21 & 13.09 & 29.82 & 9.37 & 10.20 \\
AMI-IHM & 17.01 & 19.41 & 23.63 & 10.17 & 137.15 & 129.30 & 16.93 & 17.44 \\
AMI-SDM & 41.99 & 42.64 & 42.54 & 27.99 & 257.97 & 165.16 & 37.78 & 40.29 \\
$\mu$-Bench (en-US) & 7.69 & 10.27 & 9.42 & 9.01 & 32.25 & 81.64 & 7.75 & 10.76 \\
FLEURS (en-US) & 6.61 & 6.38 & 4.96 & 4.17 & 5.58 & 13.20 & 3.88 & 3.55 \\
Earnings21 & 12.99 & 14.14 & 11.14 & 14.15 & 19.61 & 51.86 & 11.03 & 11.25 \\
\textbf{Average} & \textbf{13.59} & \textbf{13.69} & \textbf{13.56} & \textbf{10.06} & \textbf{59.98} & \textbf{61.91} & \textbf{11.40} & \textbf{12.61} \\
\midrule
\multicolumn{9}{l}{\textit{Speech Translation (BLEU, higher is better)}} \\
DE$\rightarrow$EN & --- & 46.41 & 43.65 & --- & 40.44 & 30.24 & 29.57 & 41.94 \\
FR$\rightarrow$EN & --- & 46.91 & 42.92 & --- & 40.85 & 32.30 & 34.55 & 41.77 \\
ES$\rightarrow$EN & --- & 46.96 & 44.58 & --- & 43.32 & 32.59 & 35.35 & 44.88 \\
\textbf{Translation Avg.} & --- & \textbf{46.76} & \textbf{43.72} & --- & \textbf{41.54} & \textbf{31.71} & \textbf{33.16} & \textbf{42.87} \\
\bottomrule
\end{tabular}%
}
\caption{Speech recognition and translation performance across Proprietary and Open-source systems. Recognition metrics (WER, \%) and translation metrics (BLEU) are averaged separately. Lower WER is better; higher BLEU is better. LS = LibriSpeech, $\mu$B = $\mu$-Bench (en-US), DE/FR/ES = German/French/Spanish. ASR-only models are not evaluated on Speech Translation tasks and denoted as --.}
\label{tab:speechsummary}
\end{table}

\subsection{Beyond foundational speech capabilities: Speech reasoning, instruction-following, and tool calls}

\paragraph{Recognition and multilingual translation}
\label{sec:recognitionresults}

Foundational speech capabilities---recognition and translation---establish the floor for any production voice system. \modelshort~demonstrates robust performance across these core tasks, validating that unified speech-language training supports reasoning over audio while retaining foundational speech understanding.

On LibriSpeech test-clean and test-other, \modelshort{} achieves 2.23 and 5.15 WER respectively. Across all recognition benchmarks, the average WER is 12.61, ranking among leading open-source models (\cref{tab:speechsummary}). Speech translation further validates robustness. Mean BLEU of 42.87 across German, French, and Spanish-to-English matches proprietary \gptrealtime~(43.72) and substantially outperforms other OSS systems, with Spanish-to-English reaching up to 44.88 BLEU (\cref{tab:speechsummary}).

These results establish that \modelshort{} maintains foundational speech capabilities while achieving strong reasoning performance, demonstrating the viability of unified training.

\begin{table*}[htbp]
  \centering\small
  \resizebox{\textwidth}{!}{%
  \setlength{\tabcolsep}{2pt}
  \begin{tabular}{lll|rrrrrrrr|r}
    \toprule
    Type & Source & Model & BFCL & BBA & IFEval & MTB & SS & AA & Openhermes & GSM8K & \textbf{Avg.} \\
    \midrule
    \multirow{4}{*}{3-stage} & Prop. & \fluxgpt & 71.12 & 96.8 & 53.91 & 77 & 68.88 & 68.6 & 71.6 & 84.53 & \textbf{74.06} \\
    & Mixed & \fluxgemma  & 76.43 & 98.4 & 88.15 & 84.75 & 84.82 & 75.6 & 83.4 & 91.05 & \textbf{85.33} \\
    & Mixed & \parakeetgpt & 72.77 & 98 & 49.23 & 78.12 & 85.82 & 73.4 & 74 & 87.34 & \textbf{77.34} \\
    \rowcolor{blue!10} & OSS & \parakeetgemma & 78.35 & 98.2 & 87.48 & 88.62 & 84.96 & 77.2 & 86 & 92.19 & \textbf{86.63} \\
    \midrule
    \multirow{2}{*}{S2S} & Prop. & \geminigeneral & 86.94 & 99.5 & 89 & 92.19 & 90.66 & 82.2 & 88.6 & 96.89 & \textbf{90.75} \\
    & Prop. & \gptrealtime & 80.24 & 94.6 & 88.49 & 78.44 & 85.48 & 78.4 & 84.2 & 93.18 & \textbf{85.38} \\
    \midrule
    \rowcolor{blue!10} 2-stage & OSS & \ultravoxqwen & 77.5 & 89.2 & 78.9 & 89.44 & 80.7 & 81.8 & 87.2 & 91.36 & \textbf{84.51} \\
    \rowcolor{blue!10} 2-stage & OSS & \gemmafour & 42.72 & 94 & 81.96 & 83.94 & 70.94 & 73.6 & 80 & 77.71 & \textbf{75.61} \\
    \rowcolor{blue!10} 2-stage & OSS & \audex & 55.4 & 88.59 & 71.39 & 83.19 & 70.38 & 72 & 76.4 & 91.89 & \textbf{76.16} \\
    \rowcolor{green!20} \textbf{2-stage} & \textbf{OSS} & \textbf{\model{} (thinking)} & \textbf{81.13} & \textbf{98.90} & \textbf{86.47} & \textbf{86.19} & \textbf{82.24} & \textbf{76.8} & \textbf{81.8} & \textbf{93.18} & \textbf{85.84} \\
    \bottomrule
  \end{tabular}%
  }
    \caption{\textbf{Comprehensive speech reasoning evaluation across model types and architectures.} All scores use a 0--100 scale; higher is better. \textbf{\model{} is the highest-performing open-source 2-stage model} (highlighted in green), outperforming other OSS baselines (highlighted in blue). Metrics: BFCL (Berkeley Function Calling Leaderboard), BBA (Big-Bench-Audio), IFEval (Instruction Following Evaluation), MTB (MT-Bench), SS (SpokenSquad), AA (AlpacaAudio), GSM8K (grade-school math). Type: 2-stage (speech projector + backbone), 3-stage (ASR + LLM + TTS cascade), S2S (end-to-end speech-to-speech). Source: Prop. (proprietary), Mixed (proprietary + open-source components), OSS (open-source). All models are evaluated in thinking mode and follow the individually suggested hyperparameter settings.}
    \label{tab:speechcomparison}
\end{table*}

\paragraph{Speech reasoning, instruction following, and tool calls}
In thinking mode, \model{} achieves 98.90 on Big Bench Audio and 81.13 on BFCL, closing the performance gap with proprietary systems on critical reasoning and tool-execution tasks. An overall average performance of 85.84 positions \model{} as the strongest open-source 2-stage speech-language model, substantially outperforming OSS baselines while remaining competitive with proprietary and mixed systems.

Audio-input IFEval and MT-Bench scores reach 86.47 and 86.19, respectively. The corresponding OSS baseline scores are 81.96 and 83.94 for \gemmafour~and 78.9 and 89.44 for \ultravoxqwen. \modelshort~exceeds \gemmafour~by 4.51 and 2.25 points on the 0--100 scales, while remaining competitive with \ultravoxqwen~on MT-Bench and maintaining strong performance on IFEval (\cref{tab:speechcomparison}). These results demonstrate competitive instruction-following and conversational reasoning capabilities across both benchmarks.

\begin{table}[htbp]
  \centering\small
  \setlength{\tabcolsep}{1.5pt}
  \begin{tabular*}{\columnwidth}{@{\extracolsep{\fill}}lllrrrr@{}}
    \toprule
    Type & Source & System & \multicolumn{2}{c}{EVA} & \multicolumn{2}{c}{$\tau$-Voice} \\
    & & & EVA-A & EVA-X & pass@1 & Latency (s) \\
    \midrule
    S2S & Prop. & \gptrealtime & 59.03 & 61.6 & 69.82 & 1.55 \\
    S2S & Prop. & \geminiss & 54.82 & 73.5 & 84.19 &  1.50 \\
    \midrule
    3-stage & Mixed & \fluxgemma & 36.07 & 29.22 & 65.92 & 2.92 \\
     \rowcolor{blue!10} 3-stage & OSS & \parakeetgemma & 39.97 & 3.93 & 67.78 & 3.01 \\
    \midrule
    2-stage & OSS & \ultravoxqwen & 8.94 & 1.91 & 26.80 & 3.95 \\
     \rowcolor{blue!10} 2-stage & OSS & \gemmafour &  13.25 &  5.97 & 24.20 &  3.61 \\
     \rowcolor{blue!10} 2-stage & OSS & \audex & 5.66 & 8.2 & 17.15 & 2.83 \\
    \midrule
    \rowcolor{green!20} \textbf{2-stage} & \textbf{OSS} & \textbf{\model} & \textbf{51.27} & \textbf{74.13} & \textbf{65.51} & \textbf{1.91} \\
    \bottomrule
  \end{tabular*}
  \caption{\textbf{Service accuracy and conversational experience.} EVA-A/X and $\tau$-Voice pass@1 scores use a 0--100 scale; latency is in seconds. System types: S2S (end-to-end speech-to-speech), 3-stage (ASR + LLM + TTS cascade), 2-stage (speech projector + backbone). Prop./OSS = Proprietary/Open-source. All 3-stage and 2-stage models use Cartesia Sonic 3.5 TTS~\citep{sonic35} by default. Domain-specific results are provided in Appendix \ref{sec:evataubreakdown}.}
  \label{tab:evacomparison}
\end{table}

\begin{figure}[htbp]
  \centering
  \includegraphics[trim={0.5cm 1cm 1.5cm 0.0cm}, clip, width=0.9\linewidth]{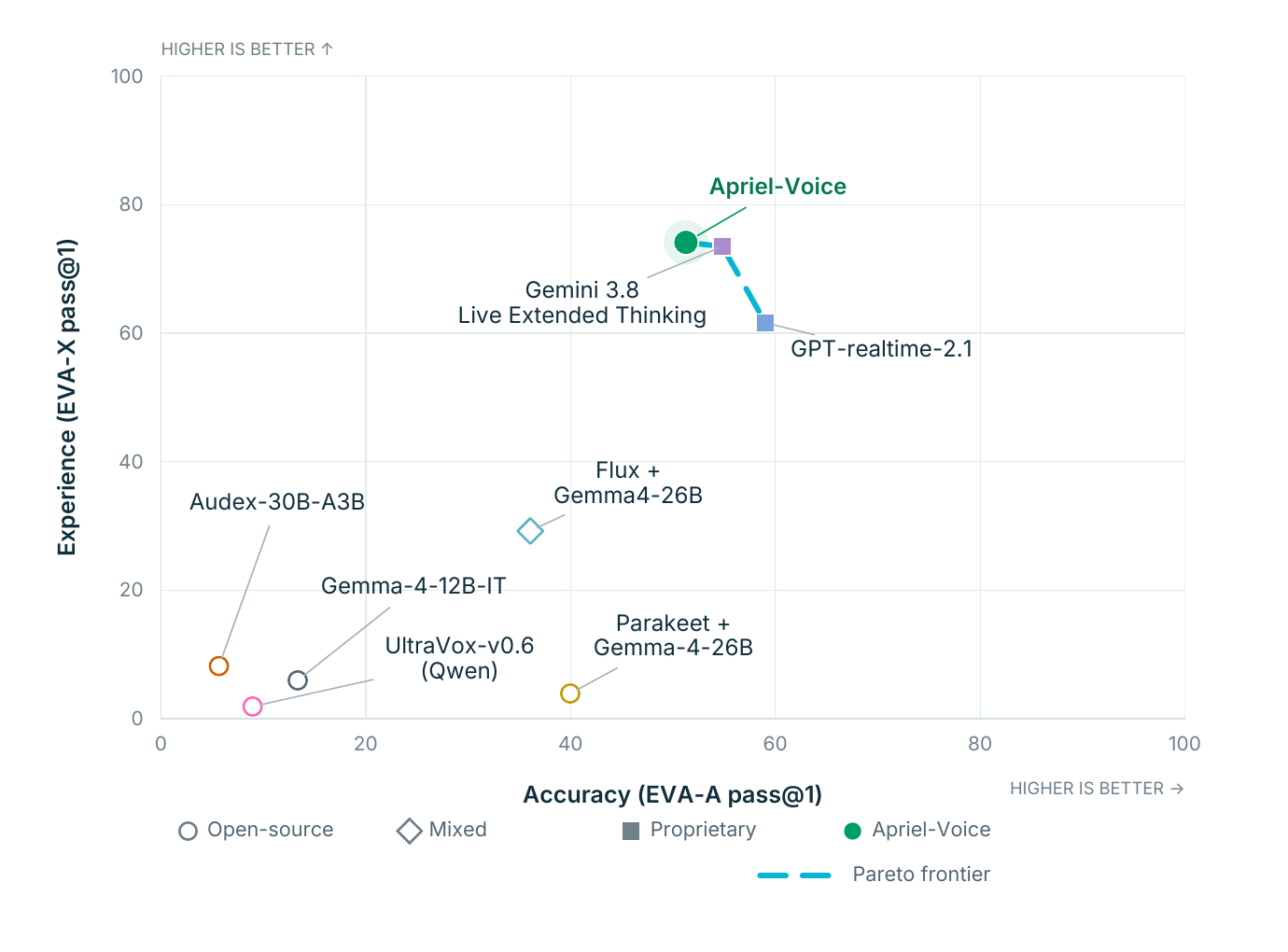}
\caption{\textbf{EVA accuracy--experience trade-offs.}
EVA-A measures task accuracy on the x-axis, while EVA-X measures conversational experience on the y-axis; higher values indicate better performance on both metrics. Marker shapes denote model source category, including open-source, mixed, proprietary, and \modelshort{}. \textbf{\modelshort{}} achieves the highest EVA-X score among the compared models while maintaining competitive EVA-A accuracy, emphasizing conversational quality without leaving the accuracy frontier. The dashed curve traces the Pareto frontier of models that are not dominated in the EVA-A--EVA-X trade-off.}
  \label{fig:eva_scatterplot}
\end{figure}

\begin{figure}[htbp]
  \centering
  \includegraphics[trim={1.0cm 1cm 1.5cm 0.0cm}, clip, width=0.9\linewidth]{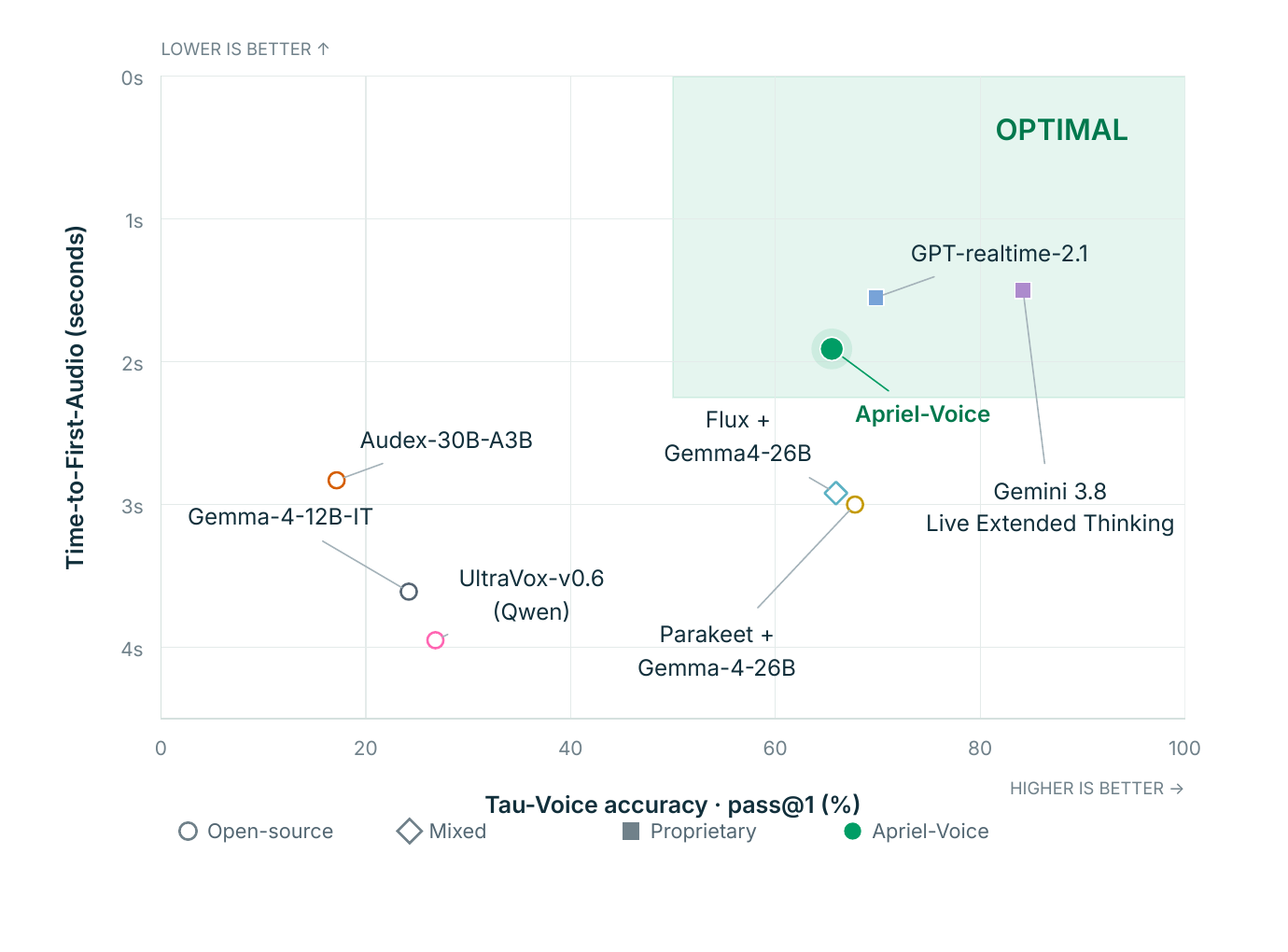}
\caption{\textbf{\tauvoice~accuracy--latency trade-offs.}
\tauvoice~pass@1 accuracy is shown on the x-axis, while Time-to-First-Audio (TTFA) is shown on the y-axis; higher accuracy and lower latency indicate better performance. Marker shapes denote model source category, including open-source, mixed, proprietary, and \modelshort{}. The shaded upper-right region highlights the preferred operating regime of high accuracy and low TTFA. \textbf{\modelshort{}} lies in the shaded optimal region, outperforming most open-source baselines.}
  \label{fig:tau_scatterplot}
\end{figure}

\subsection{Conversational Capabilities and Task Completion}
\paragraph{EVA: service accuracy and conversational experience}
With Cartesia Sonic 3.5 synthesis~\citep{sonic35}, \modelshort~reaches 51.27\% EVA-A and 74.13\% EVA-X, substantially outperforming the \parakeetgemma~baseline (39.97\% and 3.93\%) and closing the gap with proprietary systems (\geminiss~at 54.82\% and 73.5\%). Figure \ref{fig:eva_scatterplot} illustrates this positioning further: \modelshort~occupies the upper-right region alongside proprietary systems (\geminiss, \gptrealtime), while open-source baselines cluster in the lower-left. This observation establishes \modelshort~as the leading OSS model on conversational accuracy and experience in EVA. Evaluation covers 639 cases across three domains: 150 airline, 240 information technology service management (ITSM), and 249 healthcare human resources (HR). The headline values represent equal-domain means.

\begin{table}[htbp]
  \centering\small
  \setlength{\tabcolsep}{2.5pt}
  \begin{tabular*}{\linewidth}{@{\extracolsep{\fill}}lcc@{}}
    \toprule
    Component & \shortstack{Parakeet \\ + Gemma4-26B} & \shortstack{\modelshort{}} \\
    \midrule
    Speech fidelity &  \textbf{99.78} & 98.89 \\
    Faithfulness &   \textbf{66.33} & 43.49 \\
    Task completion &  60.0 & \textbf{69.94} \\
    Conciseness &  84.99 & \textbf{86.80} \\
    Conversation progression & \textbf{82.67} & 67.76 \\
    Turn taking &  40.54 & \textbf{86.52} \\
    \bottomrule
  \end{tabular*}
  \caption{\textbf{EVA component means.} \modelshort~uses thinking mode. Component scores (\%) are averaged equally across domains. EVA-A and EVA-X require their respective component criteria to pass jointly.}
  \label{tab:evacomponents}
\end{table}

Turn-taking reaches 86.52\%, compared with 40.54\% for \parakeetgemma~as a reference; and conciseness reaches 86.80\% versus 84.99\% (\cref{tab:evacomponents}).
These scores reveal \modelshort's strength in natural conversation pacing and response brevity, which are critical for user experience in deployed voice agents.
EVA-X measures whether all experience criteria pass jointly.

Performance varies across the service domains. \modelshort's airline accuracy is 59.3\%, above the reference of 54.00\%, while airline experience is 66.7\% versus 2.7\%. \modelshort~records 48.7\% accuracy and 85\% experience in ITSM, and 45.8\% and 70.7\% in healthcare HR (\cref{tab:evadomaincomparison}).

Speech fidelity is 98.89\%, while faithfulness is 43.49\%. Speech fidelity measures whether the spoken output conveys the intended response; faithfulness measures whether that response is consistent with the service interaction.

\paragraph{\tauvoice: Task completion and response latency}
\modelshort~ achieves 65.51\% mean task success on \tauvoice~in thinking mode. Airline, retail, and telecom success are 69.33\%, 81.58\%, and 45.61\%, with mean latencies of 2.22, 1.79 and 1.73 seconds (\cref{tab:taudomaincomparison}). The equal-domain latency mean is 1.91 seconds (\cref{tab:evacomparison}).

Compared to reference systems, \model{} substantially outperforms the evaluated open-source 2-stage baselines while approaching proprietary systems on \tauvoice. \geminiss~achieves the highest \tauvoice~success at 84.19\% with 1.50 seconds latency; \gptrealtime~achieves 69.82\% success at 1.55 seconds. Among open-source 2-stage baselines, \ultravoxqwen~reaches 26.80\% success at 3.95 seconds, while \gemmafour~and \audex~achieve 24.20\% and 17.15\% respectively with latencies of 3.61 and 2.83 seconds. \model{} achieves 65.51\% success at 1.91 seconds, outperforming all evaluated open-source 2-stage baselines in both task success and latency, and approaching \gptrealtime~in task success. The Gemma4-based 3-stage cascades achieve 65.92\% task success at 2.92 seconds for \fluxgemma~and 67.78\% at 3.01 seconds for \parakeetgemma. In comparison, \model{} achieves 65.51\% task success at 1.91 seconds, offering comparable task completion with substantially lower response latency.

\Cref{fig:tau_scatterplot} underscores \model{}'s competitive positioning on the task accuracy-latency trade-off. \model{} achieves substantially higher task-success rates and lower latency than the evaluated open-source 2-stage baselines, while narrowing the task-success gap with proprietary systems, demonstrating its potential for real-world voice agent applications.

\subsection{Text and Image Capability after adaptation}
\paragraph{Text-agent and enterprise workflow completion}
\begin{table}[htbp]
  \centering\small
  \setlength{\tabcolsep}{2pt}
  \begin{tabular*}{\linewidth}{@{\extracolsep{\fill}}lrr@{}}
    \toprule
    Benchmark & \shortstack{Gemma 4\\26B-A4B (T)} & Apriel Voice (T) \\
    \midrule
    MMLU-Pro & 85.27 & 84.42 \\
    GPQA Diamond~\citep{gpqa} & 77.80 & 80.40 \\
    AIME 2024 & 92.00 & 92.00 \\
    AIME 2025 & 81.67 & 81.67 \\
    MATH-500 & 92.76 & 92.88 \\
    AMC 2023 & 100.00 & 100.00 \\
    $\tau^2$ (GPT-5.2 user simulator) & 68.93 & 70.95 \\
    \bottomrule
  \end{tabular*}
  \caption{\textbf{Internal text benchmarks: Gemma 4 and Apriel Voice.} All results use text input in thinking (T) mode unless noted. The $\tau^2$ mean covers three domains (\cref{tab:tau2internal}).}
  \label{tab:textcomparison}
\end{table}

\begin{table}[htbp]
  \centering\small
  \setlength{\tabcolsep}{2.5pt}
  \begin{tabular*}{\linewidth}{@{\extracolsep{\fill}}lrrr@{}}
    \toprule
    Tools & \shortstack{Gemma 4\\26B-A4B (T)} & \shortstack{Apriel Voice\\27B Preview (T)} & \shortstack{Apriel Voice\\27B Preview (NT)} \\
    \midrule
    Oracle & 22.7 / 57.8 & 23.0 / 57.6 & 14.7 / 44.7 \\
    +5 tools & 23.4 / 60.2 & 24.7 / 61.1 & 16.0 / 46.2 \\
    +10 tools & 23.5 / 60.0 & 23.6 / 59.9 & 15.9 / 44.9 \\
    +15 tools & 25.9 / 60.9 & 23.7 / 59.4 & 14.2 / 44.8 \\
    \bottomrule
  \end{tabular*}
  \caption{\textbf{EnterpriseOps-Gym internal comparison~\citep{enterpriseops}.} Each cell gives task / verifier success (\%; $\uparrow$). Extra tools are distractors. T/NT: thinking/non-thinking. Public reference tables appear in \cref{tab:entopsaa,tab:entopsrepo}.}
  \label{tab:entopscomparison}
\end{table}

On $\tau^2$ with GPT-5.2 as the user simulator, \modelshort~ scores 70.95\% with thinking and 49.14\% without thinking; the Gemma reference scores 68.93\% and 53.23\% (\cref{tab:tau2internal}). For a higher-is-better task score $s_b$ on benchmark $b$, define the score difference
\begin{equation}
 \Delta_b=s_b^{\mathrm{Apriel\ Voice}}-s_b^{\mathrm{Gemma}}.
 \label{eq:retentiondelta}
\end{equation}
The changes are $+2.02$ and $-4.09$ percentage points, respectively. This internal comparison is distinct from the published Gemma $\tau^2$ values in \cref{tab:gemma_published}.

In the EnterpriseOps-Gym oracle setting, \modelshort~ achieves 23.0\% task success and 57.6\% verifier success with thinking, compared with 22.7\% and 57.8\% for Gemma. Task success requires every final-state check to pass; verifier success also credits partial completion.

With 5, 10, and 15 additional tools, \modelshort~ in thinking mode achieves task/verifier success rates of 24.7/61.1\%, 23.6/59.9\%, and 23.7/59.4\%. Non-thinking task success ranges from 14.2\% to 16.0\% across all four conditions (\cref{tab:entopscomparison}). Thinking has higher task success in all four conditions, while scores vary non-monotonically with the number of distractor tools. \Cref{tab:entopsaa} places the internal \modelshort~ and Gemma oracle task-success rates alongside AA's public evaluations. Both require every verifier to pass; the table identifies the different evaluation harnesses. Official repository results provide further context in \cref{tab:entopsrepo}.

\paragraph{Text and image capabilities after adaptation}
The six text-input benchmarks yield mean thinking-mode scores of 84.42 on MMLU-Pro, 80.40 on GPQA Diamond, 92.00 on AIME 2024, 81.67 on AIME 2025, 92.88 on MATH-500, and 100.00 on AMC 2023 (\cref{tab:textcomparison}).

\begin{table}[htbp]
  \centering\small
  \setlength{\tabcolsep}{2.5pt}
  \begin{tabular*}{\linewidth}{@{\extracolsep{\fill}}lrr@{}}
    \toprule
    Benchmark & \shortstack{Gemma 4\\26B} & \shortstack{Apriel Voice\\27B Preview} \\
    \midrule
    MM-IFEval~\citep{mmifeval} & \textbf{77.13} & 74.75 \\
    OCRBench~\citep{ocrbench} & \textbf{859} & 836 \\
    OCRBench v2 (EN, M)~\citep{ocrbenchv2} & \textbf{63.92} & 62.91 \\
    MathVista (M)~\citep{mathvista} & \textbf{80.50} & 79.60 \\
    MathVision (M)~\citep{mathvision} & 64.14 & \textbf{67.10} \\
    MMMU-Pro~\citep{mmmupro} & \textbf{70.23} & 69.88 \\
    CharXiv (reasoning)~\citep{charxiv} & 46.80 & \textbf{48.90} \\
    ChartQAPro~\citep{chartqapro} & 63.21 & \textbf{63.68} \\
    HallusionBench~\citep{hallusionbench} & \textbf{64.90} & 62.29 \\
    \bottomrule
  \end{tabular*}
  \caption{\textbf{Image capabilities after speech adaptation.} Internal evaluations; $\uparrow$ for all displayed scores. EN denotes English; M denotes the MINI subset. OCRBench is scored out of 1000; the other rows use a 0--100 scale. The Gemma baseline is Gemma 4 26B-A4B; both Gemma and Apriel Voice use thinking mode. Document-extraction results appear in \cref{tab:visiondocuments}.}
    \label{tab:vision}
\end{table}

On images, MMMU-Pro changes from 70.23 for Gemma to 69.88 for \modelshort, while MathVision MINI increases from 64.14 to 67.10 (\cref{tab:vision}). Three of the nine \modelshort~ scores are higher and six are lower than the Gemma reference. Document-level results appear in \cref{tab:visiondocuments}.

\section{Discussion}
\label{sec:discussion}
\subsection{Speech capability across tasks}
Apriel Voice can reason over spoken requests, follow instructions, produce tool calls, and participate in task-oriented conversations, while continuing to process text and images.

The benchmarks measure complementary capabilities. BBA and IFEval characterize different reasoning demands; BFCL tests structured calls; EVA combines service outcomes with conversational experience; and $\tau$-Voice tests completion through speech. Thinking improves several reasoning and workflow scores, while some recognition and function-calling tasks score higher in non-thinking mode. Reasoning-mode selection therefore depends on the task.

\subsection{Training and supervision}
Interface training learns a projector that connects the frozen speech encoder to the fixed decoder. Task SFT then adapts the projector and backbone for conversation and tool use. Source-dependent teacher supervision supports adaptation on selected speech examples and retention on text replay. Paired text/audio dialogues provide corresponding speech and text contexts for training.

The systems implementation addresses the resulting cost: local teacher workers separate inference from student state, lookahead overlaps teacher and student computation, and fused losses reduce temporary probability storage. Together, these mechanisms support full-vocabulary online supervision at the training scale described in \cref{tab:training}. The evaluations assess the combined method; component effects require controlled ablations.

\subsection{Optimization behavior}
\begin{figure}[htbp]
  \centering
  \includegraphics[width=0.8\linewidth]{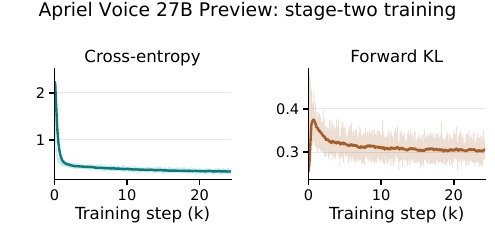}
  \caption{\textbf{Stage-two optimization.} Cross-entropy and forward KL use separate scales. Faint curves show measured losses and solid curves show smoothed trends. Components need not decrease monotonically under mixed supervision. \Cref{fig:lossfull} includes the combined objective.}
  \label{fig:loss}
\end{figure}

The combined stage-two objective decreases over training (\cref{fig:loss,fig:lossfull}). CE and FKL need not move together: CE measures fit to target responses, whereas FKL measures divergence from the frozen teacher on selected contexts. Source composition and supervision weights also affect the aggregate.

\FloatBarrier
\section{Conclusion}
\model{} extends a text-and-image backbone to speech, using the same model for reasoning, conversation, and tool-call decisions. Evaluations demonstrate these capabilities through spoken reasoning, instruction following, multilingual translation, service interactions, and complete workflows.

The recipe combines interface learning, broad task SFT, synthetic speech-conditioned dialogues, and source-dependent CE/FKL supervision. Local teacher serving, lookahead, and fused losses support online distillation. A shared real-time API enables deployment and supports integration with voice benchmarks through WebSocket communication.
\label{maintext:end}

\section*{Limitations}

\paragraph{Multilinguality.}
\model{} prioritizes English speech understanding, reasoning, and instruction following for real-world English conversational-agent settings.
Multilingual capabilities and evaluations remain limited, with primary assessment through automatic speech translation. English performance does not imply comparable quality across languages; multilingual robustness is uneven and can represent a clear limitation. Enhancing multilingual speech understanding and conversational robustness across diverse language families is a priority for future versions.

\paragraph{Evaluation and comparison.}
Internal and public results differ in prompting, language coverage, output modality, reasoning budgets, judging, and task limits. Aggregate scores do not establish paired significance, non-inferiority to the starting backbone, or a controlled ranking against every frontier system. Recognition means describe the evaluated language subsets; public and internal scoring can differ even when the language selection is shared.

\paragraph{Data and attribution.}
Synthetic generation and distillation can propagate incorrect targets or teacher predictions. Conversion-fidelity checks can miss errors and do not establish coverage of natural speech variation. The evaluations characterize the combined recipe, with no causal attribution to individual data or supervision choices. Frozen visual components also do not guarantee unchanged image reasoning through an updated decoder. Transfer to other backbone families remains untested.

\par\addvspace{2ex}
\noindent\begin{minipage}{\linewidth}
\section{Contributors}
\label{sec:contributors}

\begingroup
\raggedright
\setlength{\parskip}{4pt}
\textbf{Core Contributors}\par\nobreak
\mbox{Aman Tiwari}, \mbox{Akshay Kalkunte}, \mbox{Akintunde Oladipo}, \mbox{Jash Shah}, \mbox{Tara Bogavelli}, \mbox{Hoang Nguyen}, \mbox{Nishanth Madhusudhan}, \mbox{Vishal Juneja}\par

\vspace{5pt}
\textbf{Contributors}\par\nobreak
\mbox{Shruthan Radhakrishna}, \mbox{Masoud Hashemi}, \mbox{Kelechi Ogueji},
\mbox{Khalil Slimi}, \mbox{Saloni Mittal},
\mbox{Esakkivel Esakkiraja},
\mbox{Pulkit Pattnaik}, \mbox{Shashank Maiya}, \mbox{Dhruv Jhamb}, \mbox{Sagar Davasam}, \mbox{Katrina Stankiewicz}, \mbox{Gabrielle Gauthier Melan\c{c}on}, \mbox{Fanny Riols}, \mbox{Amit Saha}, \mbox{Lakshmi Sirisha Chodisetty}, \mbox{Sathwik Tejaswi Madhusudhan}\par

\vspace{5pt}
\textbf{Leadership}\par\nobreak
\mbox{Sridhar Nemala}, \mbox{Srinivas Sunkara}\par

\vspace{5pt}
\textbf{Acknowledgments}\par\nobreak
We also thank \mbox{Segan Subramanian}, \mbox{Anil Madamala}, \mbox{Nitin Aggarwal}, \mbox{Vikas Yadav}, and \mbox{Sai Rajeswar Mudumba} for their invaluable guidance and support.\par
\endgroup
\end{minipage}\par

{\small
\bibliographystyle{abbrvnat}
\bibliography{references}
}

\FloatBarrier
\appendix
\raggedbottom
\setlength{\intextsep}{9pt plus 1pt minus 1pt}
\setlength{\textfloatsep}{10pt plus 1pt minus 1pt}
\setlength{\floatsep}{10pt plus 1pt minus 1pt}
\makeatletter
\setlength{\@fptop}{0pt}
\setlength{\@fpsep}{12pt plus 1pt minus 1pt}
\setlength{\@fpbot}{0pt plus 1fil}
\makeatother
\section{Synthetic Data Construction Details}
\label{sec:datagenappendix}
The pipeline converts selected user content to speech while preserving the dialogue structure and assistant targets. Each converted dialogue retains a corresponding text version (\cref{fig:datageneration}).

\paragraph{Selection strategies.}
Whole-turn selection synthesizes the complete user message. Instruction-only selection does the same for generative requests without reference material. Question-only selection leaves reference material in text and speaks the question. Content-focused selection speaks material requested for reading or recitation. Multi-span selection speaks several disjoint intervals while retaining intervening text. A configurable override adds whole-turn examples. Extraction validates selected source substrings before constructing the interleaved representation. Malformed selections are retried and rejected if the retry budget is exhausted.

\paragraph{Voice assignment and chunking.}
A reference utterance is assigned once per conversation and reused across its turns and chunks. Spoken-form conversion expands written expressions into pronounceable text. Sentence-based chunking respects a duration limit. Contiguous chunks may be joined; disjoint spans retain their position among text. Speaker-aware scheduling and resumable processing preserve the reference assignment during generation.

\paragraph{Conversion fidelity.}
Spoken-form judging compares normalized content with its source. Recognition-based judging compares a synthesized chunk's transcript with the intended speech. Optional turn-level faithfulness checks assess consistency across the complete converted turn.

\section{Training Details and Loss Dynamics}
\label{sec:appendix}

\begin{figure}[htbp]
  \centering
  \includegraphics[width=\textwidth]{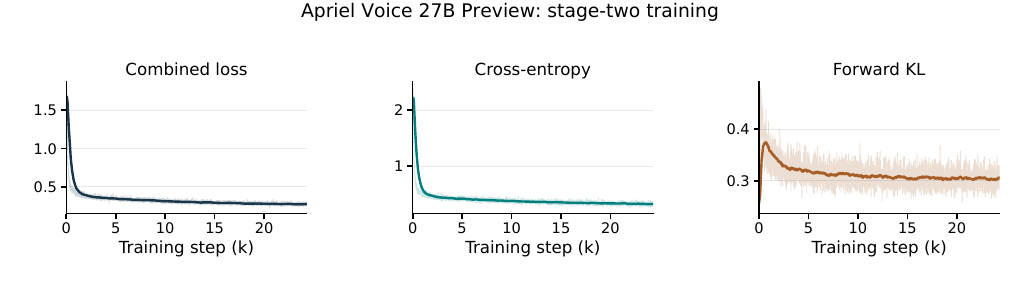}
  \caption{\textbf{Stage-two training losses.} The combined objective, CE, and FKL use separate scales. Faint curves show measured losses; solid curves show smoothed trends. The component losses reflect both the sampled tasks and their supervision weights.}
  \label{fig:lossfull}
\end{figure}

Stage one uses 1,164 dataset configurations grouped under 152 names. It contains 18.17\,M training examples, 405,931 raw bins, and 476,747 effective bins after exposure weighting, with 99.2\% token fill. Raw bins contain an average of 44.8 examples each. Per epoch, the mixture contributes approximately 1.9367\,B total tokens and 631.7583\,M loss-bearing tokens. Two completed epochs yield approximately 3.8734\,B total and 1.2635\,B loss-bearing token exposures.

The speech mixture includes Common Voice~\citep{commonvoice}, GigaSpeech~\citep{gigaspeech}, People's Speech~\citep{peoplesspeech}, LibriSpeech~\citep{librispeech}, and CoVoST~2~\citep{covost2}, with task-specific subsets and dataset variants. \Cref{tab:audiohours} distinguishes per-epoch audio duration from repeated exposure.

\begin{table}[H]
  \centering\small
  \begin{tabular*}{\linewidth}{@{\extracolsep{\fill}}lrrr@{}}
    \toprule
    Stage & Audio / epoch & Epochs & Audio exposure \\
    \midrule
    1: Interface & \stageOneAudioHours~hrs & \stageOneEpochs & \stageOneAudioExposure~hrs \\
    2: Broad SFT & \stageTwoAudioHours~hrs & \stageTwoEpochs & \stageTwoAudioExposure~hrs \\
    \bottomrule
  \end{tabular*}
  \caption{\textbf{Audio duration and training exposure.} Per-epoch durations are 32,213 and 61,540 hours. Repetition gives 64,426 and 184,620 audio-hour exposures; k denotes thousand. Text contributes tokens without an assigned audio duration.}
  \label{tab:audiohours}
\end{table}

\paragraph{Source-dependent supervision.}
Each source selects its supervision independently of its sampling weight. The pure-objective rows in \cref{tab:lossweights} are unit-weight special cases; combined CE/FKL uses separate positive coefficients, whose sum need not be one. FKL supports both selected speech tasks and retention replay.

\begin{table}[H]
  \centering\small
  \begin{tabular*}{\linewidth}{@{\extracolsep{\fill}}lcc@{}}
    \toprule
    Supervision route & CE weight & FKL weight \\
    \midrule
    Task learning: pure CE & 1 & 0 \\
    Distillation: pure FKL & 0 & 1 \\
    Weighted CE+FKL & $\alpha_d>0$ & $\beta_d>0$ \\
    \bottomrule
  \end{tabular*}
  \caption{\textbf{Source-dependent objective selection.} Internally generated responses supply task targets; the online teacher supplies full-vocabulary distributions on sources configured for distillation.}
  \label{tab:lossweights}
\end{table}

\paragraph{Packing and optimization.}
Three stage-two epochs provide approximately 52.6\,B total and 7.9\,B loss-bearing token exposures. The mixture contains 17.56\,M training examples and 1.56\,M effective bins, including exposure weighting.

Stage-two optimization uses seed 42, zero weight decay, gradient clipping at 1.0, cosine decay to 10\% of peak learning rate, and fused AdamW.

\paragraph{Loss interpretation.}
\Cref{sec:objectives} derives the objective and gradient for CE and FKL with common supervision masks and normalization. Vocabulary-wide normalization, tiled reconstruction, and token-weighted chunking are described in \cref{sec:trainingsystems}. \Cref{fig:lossfull} shows the combined objective and its CE/FKL components over the full training history.

\FloatBarrier
\section{Additional Modality Results}
\label{sec:fullmodalityresults}
Document scores (\cref{tab:visiondocuments}), and text-agent domain scores (\cref{tab:tau2internal})
complement the main capability comparisons.

\begin{table}[H]
  \centering\small
  \setlength{\tabcolsep}{2.5pt}
  \begin{tabular*}{\linewidth}{@{\extracolsep{\fill}}lrrr@{}}
    \toprule
    Benchmark & \shortstack{Gemma 4\\26B (T)} & \shortstack{Apriel Voice\\27B Preview} & GPT-4.1 \\
    \midrule
    OmniDocBench (EN)~\citep{omnidocbench} & 25.40 & 32.90 & 36.80 \\
    \bottomrule
  \end{tabular*}
  \caption{\textbf{Document-understanding results.} The OmniDocBench row covers English pages; the benchmark evaluates text, formulas, tables, and reading order~\citep{omnidocv1}. The Gemma baseline is Gemma 4 26B-A4B; both Gemma and Apriel Voice use thinking mode.}
  \label{tab:visiondocuments}
\end{table}

\begin{table}[H]
  \centering\small
  \setlength{\tabcolsep}{2.5pt}
  \begin{tabular*}{\linewidth}{@{\extracolsep{\fill}}lrrrr@{}}
    \toprule
    Agent & Airline & Retail & Telecom & Avg. \\
    \midrule
    Gemma 4 26B (T) & 72.00 & 73.10 & 61.70 & 68.93 \\
    Gemma 4 26B (NT) & 43.33 & 65.50 & 50.88 & 53.23 \\
    \midrule
    Apriel Voice 27B Preview (T) & 76.00 & 69.59 & 67.25 & 70.95 \\
    Apriel Voice 27B Preview (NT) & 38.67 & 58.77 & 50.00 & 49.14 \\
    \bottomrule
  \end{tabular*}
  \caption{\textbf{Internal $\tau^2$-Bench with GPT-5.2 as user simulator.} Task success (\%; $\uparrow$). T/NT: thinking/non-thinking. The Gemma baseline is Gemma 4 26B-A4B. Means weight the three domains equally.}
  \label{tab:tau2internal}
\end{table}

\begin{table}[H]
  \centering\small
  \setlength{\tabcolsep}{2pt}
  \begin{tabular*}{\linewidth}{@{\extracolsep{\fill}}lrrr@{}}
    \toprule
    Model & \shortstack{MMLU\\Pro} & \shortstack{GPQA\\Diamond} & $\tau^2$ Avg. \\
    \midrule
    Gemma 4 31B & 85.20 & 84.30 & 76.90 \\
    Gemma 4 26B-A4B & 82.60 & 82.30 & 68.20 \\
    Gemma 4 12B & 77.20 & 78.80 & 69.00 \\
    \bottomrule
  \end{tabular*}
  \caption{\textbf{Published Gemma 4 text benchmarks.} Officially reported results are extracted from~\citep{gemma4card}.}
  \label{tab:gemma_published}
\end{table}

\FloatBarrier
\section{Enterprise Workflow Evaluation}
\label{sec:enterprisedetails}
EnterpriseOps-Gym evaluates stateful tool use across eight enterprise domains~\citep{enterpriseops}. \Cref{tab:entopsaa} compares internal oracle task-success rates for Apriel Voice and Gemma 4 26B-A4B with independent Artificial Analysis evaluations; \cref{tab:entopsrepo} gives official repository results. Strict success requires all final-state checks to pass (\cref{eq:entops}). The internal verifier-success diagnostics and distractor-tool conditions appear in \cref{tab:entopscomparison}.

\begin{table}[H]
  \centering\small
  \setlength{\tabcolsep}{2.5pt}
  \begin{tabular*}{\linewidth}{@{\extracolsep{\fill}}lr@{}}
    \toprule
    Model / setting & Task success (\%) \\
    \midrule
    \multicolumn{2}{l}{\textit{Independent Artificial Analysis evaluations}} \\
    Claude Fable 5 (with fallback) & 51.1 \\
    Gemini 3.7 Flash (medium) & 50.4 \\
    Gemini 3.5 Flash & 50.1 \\
    DeepSeek V4 Pro 0813 (max) & 49.6 \\
    Grok 4.6 (high) & 48.3 \\
    Claude Opus 5 (max) & 47.5 \\
    Qwen3.8 2.4T A95B & 47.4 \\
    Muse Spark 1.2 (xhigh) & 47.3 \\
    DeepSeek V4 Flash Vision (max) & 47.1 \\
    GPT-5.5 (xhigh) & 46.6 \\
    Kimi K3 (max) & 45.3 \\
    Qwen3.8 27B (xhigh) & 44.2 \\
    GPT-5.6 Sol (max) & 42.9 \\
    Gemini 3.5 Flash-Lite & 42.3 \\
    GPT-5.6 Luna (max) & 40.8 \\
    GPT-5.6 Terra (max) & 38.5 \\
    Inkling & 38.0 \\
    GLM-5.3 (max) & 36.4 \\
    Muse Glimmer (high) & 34.7 \\
    Mistral Medium 3.5 & 33.7 \\
    \midrule
    \multicolumn{2}{l}{\textit{Internal evaluations; oracle tools}} \\
    Gemma 4 26B-A4B (T) & 22.7 \\
    \textbf{Apriel Voice 27B Preview (T)} & 23.0 \\
    \textbf{Apriel Voice 27B Preview (NT)} & 14.7 \\
    \bottomrule
  \end{tabular*}
  \caption{\textbf{EnterpriseOps-Gym oracle task success.} All rows report strict task success (\%; $\uparrow$) with oracle tools: every final-state verifier must pass. AA runs use Stirrup and three repeats per task~\citep{entopsaa}. Internal Gemma and Apriel Voice evaluations use a different harness and are grouped separately. T/NT: thinking/non-thinking. Verifier success and distractor-tool settings appear in \cref{tab:entopscomparison}.}
  \label{tab:entopsaa}
\end{table}

\begin{table}[H]
  \centering\small
  \setlength{\tabcolsep}{2.5pt}
  \begin{tabular*}{\linewidth}{@{\extracolsep{\fill}}lr@{}}
    \toprule
    Model / setting & Task success (\%) \\
    \midrule
    Claude Opus 4.6 & 45.9 \\
    Claude Sonnet 4.6 & 42.2 \\
    Claude Opus 4.5 & 39.4 \\
    Gemini-3.1-Pro & 38.7 \\
    Claude Sonnet 4.5 & 34.1 \\
    Gemini-3-Flash & 34.0 \\
    Gemini-3-Pro & 29.7 \\
    GPT-5 & 30.9 \\
    GPT-5-Mini & 22.9 \\
    Gemini-2.5-Pro & 20.0 \\
    DeepSeek-V3.2 & 24.2 \\
    Kimi-K2-Thinking & 22.4 \\
    Qwen3-30B (Think) & 19.1 \\
    Qwen3-235B (Inst.) & 18.1 \\
    Qwen3-4B (Think) & 15.3 \\
    \bottomrule
  \end{tabular*}
  \caption{\textbf{EnterpriseOps-Gym repository reference models.} Task success (\%; $\uparrow$) with oracle tools on the full benchmark, from the official repository~\citep{entopsrepo}. Think and Inst. denote thinking and instruction-tuned variants.}
  \label{tab:entopsrepo}
\end{table}

\FloatBarrier

\wideappendixheader{%
  \section{EVA and \texorpdfstring{\tauvoice}{Tau-Voice}~Domain Breakdown}
  \label{sec:evataubreakdown}
  \Cref{tab:evadomaincomparison,tab:taudomaincomparison} report domain-level results for EVA and \texorpdfstring{\tauvoice}{Tau-Voice}~respectively.

\begin{table}[H]
  \centering\fontsize{9}{10}\selectfont
  \setlength{\tabcolsep}{2.5pt}
  \begin{tabular*}{\textwidth}{@{\extracolsep{\fill}}llcccccccc@{}}
    \toprule
    Type & Source & System & \multicolumn{2}{c}{Airline} & \multicolumn{2}{c}{ITSM} & \multicolumn{2}{c}{HR} \\
    & & & EVA-A & EVA-X & EVA-A & EVA-X & EVA-A & EVA-X \\
    \midrule
    S2S & Prop. & \gptrealtime & 54.0 & 51.3 & 72.9 & 68.8 & 50.2 & 64.7 \\
    S2S & Prop. & \geminiss & 62.67 & 70.0 & 60.0 & 75.4 & 41.8 & 75.1 \\
    \midrule
    3-stage & Mixed & \fluxgemma & 47.3 & 30.7 & 25.83 & 28.33 & 35.08 & 28.63 \\
    \rowcolor{blue!10} 3-stage & OSS & \parakeetgemma & 54.0 & 2.7 & 40.9 & 9.1 & 25.0 & 0.0 \\
    \midrule

    \rowcolor{blue!10} 2-stage & OSS & \ultravoxqwen & 15.33 & 0.0 & 7.08 & 2.92 & 4.42 & 2.81 \\
    \rowcolor{blue!10} 2-stage & OSS & \gemmafour & 28.0 & 0.0 & 3.33 & 17.92 & 8.43 & 0.0 \\
    \rowcolor{blue!10} 2-stage & OSS & \audex & 8.0 & 12.0 & 3.75 & 4.17 & 5.22 & 8.43 \\
    \rowcolor{green!20} 2-stage & OSS & \textbf{\model{}} & \textbf{59.3} & \textbf{66.7} & \textbf{48.7} & \textbf{85.0} & \textbf{45.8} & \textbf{70.7} \\
    \bottomrule
  \end{tabular*}
  \caption{\textbf{EVA accuracy and experience by domain.} EVA-A and EVA-X scores (higher is better) across airline, ITSM (information technology service management), and healthcare HR domains. All 2-stage and 3-stage models use thinking mode and Cartesia Sonic 3.5~\citep{sonic35} for consistency and fair comparisons. OSS baselines are highlighted in blue; \model{} in green.}
  \label{tab:evadomaincomparison}
\end{table}

\begin{table}[H]
  \centering\fontsize{9}{10}\selectfont
  \setlength{\tabcolsep}{2.5pt}

  \begin{tabular*}{\textwidth}{@{\extracolsep{\fill}}lllcccccc@{}}
    \toprule
    Type & Source & System & \multicolumn{2}{c}{Airline} & \multicolumn{2}{c}{Retail} & \multicolumn{2}{c}{Telecom} \\
    & & & Pass@1 & Latency & Pass@1 & Latency & Pass@1 & Latency \\
    \midrule

    S2S & Prop. & \gptrealtime & 70.00 & 1.80 & 77.19 & 1.66 & 62.28 & 1.19 \\

    S2S & Prop. & \geminiss & 78.00 & 1.44 & 77.19 & 1.50 & 97.37 & 1.55 \\

    \midrule

    3-stage & Mixed & \fluxgemma & 65.30 & 3.24 & 79.82 & 2.87 & 52.63 & 2.64 \\

    \rowcolor{blue!10} 3-stage & OSS & \parakeetgemma & 70.00 & 3.37 & 78.95 & 3.00 & 54.39 & 2.65 \\

    \midrule
    \rowcolor{blue!10} 2-stage & OSS & \ultravoxqwen & 42.67 & 3.83 & 18.71 & 3.85 & 19.01 & 4.18 \\

    \rowcolor{blue!10} 2-stage & OSS & \gemmafour & 34.00 & 3.83 & 20.18 & 3.49 & 18.42 & 3.52 \\

    \rowcolor{blue!10} 2-stage & OSS & \audex & 42.67 & 2.33 & 6.73 & 2.70 & 2.05 & 3.45 \\

    \rowcolor{green!20} 2-stage & OSS & \textbf{\model{}} & \textbf{69.33} & \textbf{2.22} & \textbf{81.58} & \textbf{1.79} & \textbf{45.61} & \textbf{1.73} \\

    \bottomrule
  \end{tabular*}

  \caption{\textbf{Task success and latency on $\tau$-Voice by domain.} Pass@1 (\%) and latency (s) across airline, retail, and telecom domains. \model{} uses thinking mode with Cartesia Sonic 3.5~\citep{sonic35} for consistency with reference systems. OSS baselines are highlighted in blue; \model{} in green.}

  \label{tab:taudomaincomparison}
\end{table}

}

\end{document}